\documentclass[letter,reqno]{qt}

\title{Asymptotic Purification of Quantum Trajectories under Random Generalized Measurements}

\usepackage{authblk}

\author[1]{Owen Ekblad\thanks{ekbladow@msu.edu}}
\author[2]{Eloy Moreno-Nadales\thanks{morenon4@msu.edu}}
\author[3]{Lubashan Pathirana\thanks{lpk@math.ku.dk}}
\author[4]{Jeffrey Schenker\thanks{schenke6@msu.edu}}
\affil[1,2,4]{Department of Mathematics, Michigan State University.}
\affil[3]{QMATH, University of Copenhagen}
\date{}

\begin{document}

\pagenumbering{arabic}
\lhead{\thepage}
\maketitle

\vspace{-1cm}
\begin{abstract}
We develop a general framework to study quantum trajectories resulting from repeated random measurements subject to stationary noise, and generalize results of K\"ummerer and Maassen \cite{maassen2006purification} to this setting. The resulting trajectory of quantum states is a time-inhomogeneous Markov chain in a random environment. K\"ummerer and Maassen introduced the concept of dark subspaces for noise-free processes, establishing that their absence is equivalent to asymptotic purification of the system state. We clarify the notion of dark subspaces in the disordered setting by defining a measurable correspondence consisting of a collection of random subspaces satisfying a darkness condition. We further prove that asymptotic purification occurs if and only if this collection of random dark subspaces is empty. Several examples of these phenomena are provided.
\end{abstract}

%
%


%
%
 
\section{Introduction}\label{Sec1:Main}
The inevitable effect of a measurement on a quantum system has been a 
fundamental part of quantum theory since its inception 
\cite{born1926,heisenberg1927,vonneumann1932,wigner1963}. In his classic text 
\cite{vonneumann1932}, von Neumann emphasized that quantum mechanics contains 
``two fundamentally different types" of state changes: unitary evolution and 
measurement. The unitary evolution of a system is the reversible change 
generated by the Schr\"odinger equation; over a discrete unit of time it 
results in the transformation $\rho \mapsto U\rho U^\dagger$ of the density 
matrix $\rho$ describing the state of the system, with $U$ a unitary operator.  
However, the state change induced by a measurement, sometimes referred to as 
``the collapse of the wave function,'' is less clearly defined.  
Born's rule \cite{born1926} provides an unambiguous procedure for computing the probability distribution for the outcomes of a measurement.
Let $A$ be an observable with purely discrete spectrum, consisting of eigenvalues $\lambda_j$, $j=1,\ldots,N$, with corresponding eigen-projections $P_j$. Born's rule states that the probability a measurement of $A$ applied to a system in a state with density matrix $\rho$ leads to the result $\lambda_j$ is given by $\mathrm{Pr}[a=\lambda_j] = \mathrm{tr} \left ( P_{j} \rho \right ).$ However, Born's rule does not specify the state of the system after a measurement. For certain measurements, one expects the following principle to hold 
\begin{description}
    \item{\textbf{M}}. \emph{If the physical quantity $\mathcal{R}$ is measured twice in prompt succession in a system $\mathbf{S}$ then we get the same value each time.} \cite[\S IV.3]{vonneumann1932}
\end{description}
\noindent However, even this principle is not enough to specify the state following a measurement, unless the eigenprojection $P_j$ is rank one.

Already in \cite{vonneumann1932}, it was noted that a proper physical analysis requires treating part of the measuring apparatus itself as a quantum object. Physically, the measurement of a quantum system $\mcS$ is usually indirect and entails the direct measurement of an ancillary system $\mcR$ upon interaction between $\mcS$ and $\mcR$.  Careful analyses of this type by various authors in the 1960's and 1970's led to a formulation of ``generalized'' or ``indirect'' measurements;  see \cite{kraus1983states} for a review, and \cite{davies1970,hellwig1970,hellwig1969,ludwig1974} for some of the original literature.  Such measurements are often called \emph{Kraus measurements}, after Karl Kraus.

In this work, we restrict our attention to finite dimensional quantum systems and to \textit{perfect} Kraus measurements, modelled as follows. We consider a quantum system $\mcS$ with state space modelled by the set of density matrices on a finite dimensional Hilbert space $\mbC^d$.  A \emph{perfect Kraus measurement} on $\mcS$ with outcomes indexed by a finite set $\mcA$ is described by a collection $\mcV = \{V_a\}_{a\in\mcA}$ of matrices $V_a\in\mbM_{d\times d}(\mbC)$ satisfying the equation 
\begin{equation}\label{Trace preservation of Kraus measurement}
    \sum_{a\in\mcA}V_a^\dagger V_a  \ = \ I \ ,
\end{equation}
where $\dagger$ denotes the conjugate transpose and $I$ is the $d\times d$ identity matrix.  
If the system $\mcS$ in the state described by $\rho_0$ is measured according to $\mcV$, then the state following the measurement is updated according to
\begin{equation}\label{Quantum trajectory as Markov chain, one measurement}
    \rho_0 \ \mapsto \
    \frac{\Va{\rho_0}V_a^\dagger}{\Pa{\rho_0}}
    \quad\text{with probability }\Pa{\rho_0} \ .
\end{equation}

According to \cite{kraus1983states}, each physical indirect measurement apparatus $\scrA$ with outcomes $\mcA$ of a quantum system $\mcS$ corresponds to a perfect Kraus measurement $\mcV = \{V_a\}_{a\in\mcA}$ as defined above.
This formalism allows one to consider repeated measurements.  Indeed, given that the system $\mcS$ starts in state $\rho_0$, one finds that, upon measuring the system $n$-many times, the updated density matrix is given by
\begin{equation}\label{Quantum trajectory as Markov chain}
    \rho_0\ \mapsto \  
    \frac{V_{a_n}\cdots V_{a_1}\rho_0 V_{a_1}^\dagger\cdots V_{a_n}^\dagger}{\tr{V_{a_n}\cdots V_{a_1}\rho_0 V_{a_1}^\dagger\cdots V_{a_n}^\dagger}}
    \quad\text{with probability }
    \tr{V_{a_n}\cdots V_{a_1}\rho_0 V_{a_1}^\dagger\cdots V_{a_n}^\dagger} \ .
\end{equation}
Letting $\rho^{(n)}_{\rho_0}$ denote the state of $\mcS$ after $n$-many measurements, we refer to the sequence $\bigl (\rho^{(n)}_{\rho_0} \bigr)_{n\in\mbN}$ as a \textit{quantum trajectory} corresponding to Kraus measurement $\mcV$, with the initial state $\rho_0$. Note that $\rho^{(n)}_{\rho_0}$ is a function of sequences of measurement outcomes $\bar{a}\in\mcA^\mbN$ and the initial state $\rho_0$.
Kraus measurements can model indirect measurements that individually result in a very modest change to the state of the system.  For example, the operators $\{V_a\}_{a\in \mcA}$ may all be invertible, so that in principle the effect of a measurement is reversible. However, the long term effect of repeated measurements may still be irreversible and lead to a phenomenon akin to ``wave packet reduction,'' namely \emph{asymptotic purification}.  Recall that a \emph{pure state} is one with a rank-one density matrix. The main theorem of a 2006 paper by Maassen and K\"ummerer's \cite{maassen2006purification} states that either
\begin{enumerate}
    \item[(i)] \emph{there exists a so-called \emph{dark subspace} for $\mcV$, which is a subspace $W$ with $\dim W \ge 2$ such that any product $V_{a_n}\cdots V_{a_1}$ restricted to $W$ is proportional to an isometry, or}

    \item[(ii)] \emph{for all initial states the trajectory asymptotically approaches the set of  pure states, i.e., \emph{asymptotically purifies}, on the set of physically significant sequence of measurement outcomes.}
\end{enumerate}
Heuristically, a dark subspace is one from which information does not leak. 
In this work, we generalize the above result to measurement processes that occur in the presence of time-dependent classical noise, either in the environment or in the measurement apparatus itself. 
In the presence of noise, the Kraus measurement depends on the realization of the disorder, which we model as a point $\omega\in \Omega$ in a probability space $(\Omega,\mbP)$. 
We model the noisy situation by a perfect Kraus measurement-valued random variable
\begin{equation}
    \mcV: \Omega \ \to \ \{\text{perfect Kraus measurements}\}, \quad \mcV_\omega \ = \ \{V_{a;\omega}\}_{a\in\mcA} \ .
\end{equation}
We have made the natural but somewhat simplifying assumption that the set of possible measurement outcomes $\mcA$ does not depend on the disorder.\footnote{In practice, the set of actually realizable measurement outcomes may depend on $\omega$, since $V_{a;\omega}$ could vanish for some $a$ with positive probability, so long as the total set of all possible outcomes for any disorder configuration is finite.}
We are interested in time-dependent disorder.  We work in the discrete time setting, and take the transformation of the disorder over a single time step to be an ergodic, measure preserving transformation $\theta : \Omega \to \Omega$.  That is, if the disorder is in configuration $\omega_0$ at time step $n_0\in \mathbb{Z}$, then at a later time step $n_1>n_0$ the disorder is found in configuration $\theta^{n_1-n_0}(\omega_0)$.  
We thus modify \eqref{Quantum trajectory as Markov chain} to read 
\begin{multline}\label{Quantum trajectory as disordered Markov chain}
    \rho_0\ \mapsto \  
    \frac{V_{a_n; \theta^n(\omega)}\cdots V_{a_1; \theta(\omega)}\rho_0 V_{a_1; \theta(\omega)}^\dagger\cdots V_{a_n; \theta^n(\omega)}^\dagger}{\tr{V_{a_n; \theta^n(\omega)}\cdots V_{a_1; \theta(\omega)}\rho_0 V_{a_1; \theta(\omega)}^\dagger\cdots V_{a_n; \theta^n(\omega)}^\dagger}}\\
    \text{with probability }\tr{V_{a_n; \theta^n(\omega)}\cdots V_{a_1; \theta(\omega)}\rho_0 V_{a_1; \theta(\omega)}^\dagger\cdots V_{a_n; \theta^n(\omega)}^\dagger} \ , 
\end{multline}
and, in this way, we obtain a \textit{disordered quantum trajectory}, which we also denote by $\bigl(\rho^{(n)}\bigr)_{n\in\mbN}$. 
One of the main objects introduced here is a generalization of the notion of a dark subspace mentioned in (ii) above. We find that, in our setting, the appropriate notion is captured by considering a random compact set $\omega\mapsto \scrD_\omega$, consisting of all projections in $\mbM_{d\times d}(\mbC)$ of rank 2 or higher satisfying a certain ``darkness'' condition; a technical discussion of random sets is given in \S\ref{Sec:Dark subspaces}. We demonstrate that the random set $\scrD$ of dark projections classifies the asymptotic purification of the random Markov chain $(\rho^{(n)}_\omega)_{n\in\mbN}$ analogous to the non-disordered setting. Namely, we prove that either
\begin{enumerate}
    \item[(i$'$)] \emph{for almost every $\omega\in\Omega$, the set $\scrD_\omega$ is nonempty, i.e., almost surely there exists a dark subspace for $\mcV_\omega$, or} 

    \item[(ii$'$)] \emph{for almost every $\omega\in\Omega$, for all initial states the trajectory $\bigl(\rho^{(n)}_\omega(\bar{a}) \bigr)_{n\in\mbN}$ asymptotically approaches the pure states on the set of physically significant measurement outcomes.}
\end{enumerate}
Specifically, letting $X$ denote the set of $\omega\in\Omega$ such that for all initial states asymptotic purification  of the trajectory occurs on the set of physically significant measurement outcomes, we prove that $X$ is a measurable event with trivial probability (i.e. $\Pr(X)\in\{0, 1\}$), and that the equality $\Omega\setminus X = \{\scrD\neq \emptyset\}$ holds almost surely. In particular, to show that $\Pr(X) = 1$, for example, it suffices to show that $\Pr(X) > 0$, which one can do by establishing the equivalent condition that $\Pr(\scrD = \emptyset)>0$. As explained in Remark \ref{rem:constructive}, this observation provides a constructive criterion for asympotitic purification.
By means of the Ergodic Decomposition Theorem \cite[Theorem 5.1.3]{Viana_Oliveira_2016}, we may extend the above results to apply to stationary but not necessarily ergodic dynamics (assuming $\Omega$ to be a standard probability space). 
Loosely speaking, any stationary random process may be modeled as a random choice (according to some probability measure) of an ergodic process obtained by restricting the dynamics to a measurable subset of $\Omega$. See Remark \ref{Rmk:Ergodic decomposition for stationary} below for more details.
\subsection{Background and relation to other work}
As outlined above, our work may be viewed as a generalization of results of K\"ummerer and Maassen \cite{maassen2006purification} to a setting incorporating time-dependent disorder. As in \cite{maassen2006purification}, we work in the discrete time setting. Within the physics literature a \emph{quantum trajectory} is often modelled as the \textit{continuous}-time process $(\Theta_t)_{t\geq 0}$ satisfying stochastic Schr\"odinger equations; see \cite{barchielli2009quantum, bouten2004stochastic, carmichael1991anopen, davies1976open, gisin1984quantum, pellegrini2008existence, wiseman2009quantum}.  The convergence of discrete-time models of quantum trajectories to a continuous limit is well understood; see \cite{bauer2013repeated,pellegrini2008existence}.
The formalism of quantum trajectories was used to describe the Nobel-prize winning experiment of Haroche \cite{guerlin2007progressive}.  The interested reader may consult any of \cite{ barchielli2009quantum, holevo2003statistical, wiseman2009quantum} for an exposition of quantum trajectories and \cite{carmichael1991anopen} for an exposition of quantum optics, but we shall not comment further on the general physical theory of quantum trajectories. 
The concept of a dark subspace was introduced in \cite{maassen2006purification}. These were shown to be related to quantum error-correction in \cite{majgier2010protected}. In \cite{benoist2019invariant}, it was shown that, under an irreducibility assumption, quantum trajectories with no dark subspaces necessarily have a unique invariant measure. Further asymptotic behaviour for such trajectories, including a law of large numbers and a functional central limit theorem, are  studied in \cite{benoist2023limit}. 
Quantum trajectories perturbed by certain types of disorder have been studied previously. In \cite{bauer2015computing}, the stochastic differential equations describing a quantum trajectory were studied under the assumption of external Gaussian noise. In \cite{ballesteros2019perturbation}, the perturbation theory of a certain type of quantum trajectory was studied, and the asymptotic behavior was studied in detail. The current model of noise was studied by Schenker and Movassagh in \cite{movassagh2022ergodic, movassagh2021theory} and further by Schenker and Pathirana in \cite{pathirana2023law}.  
Repeated quantum measurements have also been studied from the perspective of nonequilibrium statistical mechanics.  In \cite{benoist2021b, benoist2018}, the authors derive a large deviation principle for the results of repeated measurements and use it give a physically meaningful definition of the entropy production in such systems.  These results were extended in \cite{bougron2022} to a model incorporating Markovian disorder.  In forthcoming work with Renaud Raqu\'epas, one of us has extended these results to the ergodic context considered here \cite{raquepas2024}.

By modelling disorder with an ergodic dynamical system, we incorporate many different models of disorder into a single framework. For example, an ergodic dynamical system can model an i.i.d. sequence of random variables, a stationary Markov chain, some sequences of random variables with long-range correlations, and even quasi-periodic dynamics.  
\subsection{Organization}
In \S \ref{Sec:Disordered quantum trajectories}, we formally introduce the main mathematical objects of study and state our theorems. We also define the concept of \emph{physically significant sequence of outcomes},  a probability measure on the space of sequences of observed outcomes that depends on the initial state of the system. The bulk of the results that are needed to prove  Theorems \ref{0-1 law for purification of disordered quantum trajectories}, \ref{Main theorem} and \ref{Main theorem'} are presented in \S \ref{Sec:Purification event}, where we also formally introduce the \emph{purification event} and prove Theorem \ref{0-1 law for purification of disordered quantum trajectories}.  In \S \ref{Sec:Dark subspaces}, we introduce the notions of a \emph{random gray projections} and \emph{random dark projections} as random set-valued functions. In \S\ref{Sec:Dark subspaces} we also provide the proof of theorem \ref{Main theorem} and \ref{Main theorem'}. Two technical results that play a key role in the proof of Theorem \ref{0-1 law for purification of disordered quantum trajectories} are presented in an Appendix.
 
%
%

\section{Disordered quantum trajectories and dark subspaces}\label{Sec:Disordered quantum trajectories}
\subsection{Notation}
Let $\matrices$ denote the space of $d\times d$ matrices with complex-valued entries. For $M\in\matrices$, we let $M^\dagger$ denote the conjugate transpose of $M$, and write $|M|$ to denote the unique square root of the matrix $M^\dagger M$. We view $\matrices$ as a finite-dimensional Banach space with the trace norm $\|M\|_1 = \operatorname{Tr}|M|$.  Given a Borel set $S \subset \mcB(S)$, we denote the $\sigma$-algebra of Borel subsets of $S$ by $\mcB(S)$.
We denote the closed cone of positive semi-definite matrices in $\matrices$ by
\begin{equation}
        \mbP_d
            \ := \ 
        \left\{
            M\in\matrices
                \,:\,
            M = |M|
        \right\} \ .
\end{equation}
If $M$ belongs to $\mbP_d$, we write $M\geq 0$. We denote the set of states, i.e., density matrices, by
\begin{equation}
    \mbS_d
        \ := \ 
    \left\{ 
        \rho\in\matrices 
            \,:\,
        \text{$\rho\geq 0$ and $\tr{\rho} = 1$}
    \right\} \ .
\end{equation}
We say that $p\in\matrices$ is a projection if $p^2 = p = p^\dagger$. The rank of a projection $p$ is defined to be the dimension of its range, which is equal to $\tr{p}$. We denote by $\scrP$ the set of all projections in $\matrices$, and by $\scrP_r$ we mean the set of projections of rank $r$, 
\begin{equation}
    \scrP_r
        \ = \ 
    \left\{
        p\in\scrP 
            \,:\,
        \text{$\operatorname{tr}(p) = r$}
    \right\}\ .
\end{equation}
Let $\mcA$ be a finite set. We let $\scrV_\mcA$ denote the set 
\begin{equation}\label{Definition of the set of perfect Kraus measurements on A}
    \scrV_\mcA 
            \ := \ 
        \left\{
            \left\{V_a\right\}_{a\in\mcA}\subset\matrices
                \,\,:\,\,
            \sum_{a\in\mcA}V_a^\dagger V_a = \mbI
        \right\} \ .
\end{equation}
An element $\mcV$ of $\scrV_\mcA$ is called a \textit{perfect Kraus measurement} and $\mcA$ the set of observable outcomes. Note that $\mcV\in\scrV_\mcA$ if and only if the linear map 
\begin{equation}\label{Trace preservation for perfect Kraus measurement}
    \begin{split}
        \Phi_\mcV: \matrices &\to \matrices \ , \\
        M &\ \mapsto \   \sum_{a\in\mcA}V_a M V_a^\dagger 
    \end{split}
\end{equation}
satisfies $\tr{\Phi_\mcV(M)} = \tr{M}$ for all $M\in\matrices$.
Let $(\Omega, \mcF,\Pr)$ be a probability space together with a measurably invertible, measure-preserving, ergodic transformation $\theta : \Omega\to\Omega$. Recall that a measure preserving transformation (m.p.t.) $\theta$ is called \textit{measurably invertible} (or simply \textit{invertible}) if $\theta$ is a measurable bijection with measurable inverse, and that $\theta$ is \textit{ergodic} if every $\theta$-invariant measurable set has probability $0$ or $1$. That is, $\theta$ is ergodic if $\theta$ is measure-preserving and for any measurable set $A\in\mcF$ such that $\theta^{-1}(A) = A$, it holds that $\Pr(A)\in\{0, 1\}$.
For a m.p.t. $\theta$, ergodicity is in fact equivalent to the seemingly stronger property that essentially $\theta$-invariant measurable sets have probability $0$ or $1$, i.e., for $A\in\mcF$, if $\Pr\big(\theta^{-1}(A)\Delta A\big) = 0$, then $\Pr(A)\in\{0, 1\}$ \textemdash \ see \cite[Theorem 1.5]{walters2000introduction}. 
\begin{remark}
Not every ergodic m.p.t.\ is invertible.  However, for a large class of probability spaces (namely, Lebesgue probability spaces), any \emph{surjective} ergodic m.p.t.\ can be extended to an ergodic and invertible m.p.t.\ on larger space in the following manner. Let $\theta:\Omega\to\Omega$ be a surjective m.p.t. defined on  $(\Omega,\mcF,\Pr)$. Then there exists a probability space $(\tilde{\Omega}, \tilde{\mcF}, \tilde{\Pr})$, an invertible, measure-preserving, ergodic map $\tilde{\theta}:\tilde{\Omega}\to\tilde{\Omega}$, and an injective measure-preserving map $\iota:\Omega\to\tilde{\Omega}$ with measure-preserving (and measurable) inverse $\iota^{-1}:\iota(\Omega)\to\Omega$ such that the diagram
\begin{equation}
    \begin{tikzcd}
	\Omega && \Omega \\
	{\tilde{\Omega}} && {\tilde{\Omega}}
	\arrow["\theta", from=1-1, to=1-3]
	\arrow["{\tilde{\theta}}", from=2-1, to=2-3]
	\arrow["\iota"', from=1-1, to=2-1]
	\arrow["\iota", from=1-3, to=2-3]
\end{tikzcd}
\end{equation}
commutes; see \cite{rohlin1964exact, rohlin1961exact} or \cite[\S10]{cornfeld2012ergodic}. Such a construction is known as the \textit{natural extension} of $\theta$. In particular, for many common examples of measure-preserving and ergodic maps --- such as the one-sided shift on a sequence space --- such an invertible extension exists. 
\end{remark}
Let $\mcV = \{V_a\}_{a\in\mcA}$ be a collection of random matrices $V_a: \Omega \to \matrices$ satisfying $\mcV_\omega\in\scrV_\mcA$ for almost every $\omega\in\Omega$. Whenever possible, we suppress the independent variable $\omega\in\Omega$ in formulas involving random variables. When necessary, however, we denote the dependence of a random variable on $\omega\in\Omega$ by a subscript. For example, $\mcV\in\scrV_\mcA$ almost surely expresses that 
\begin{equation}\label{krausmeas}
    \sum_{a\in\mcA}V_{a;\omega}^\dagger V_{a;\omega} \ = \  I
\end{equation}
holds for almost every $\omega\in\Omega$. 
\begin{definition}\label{Definition of disordered perfect Kraus measurement}
Given a finite set of outcomes $\mcA$, a collection $\mcV = \left\{V_{a}\right\}_{a\in\mcA}$ of random matrices $V_a$ such that $\mcV\in\scrV_\mcA$ almost surely is called a \emph{disordered perfect Kraus measurement} associated to $\mcA$. 
\end{definition}
Consider now the set $\mcA^\mbN$ of sequences of measurement outcomes. We write $\bar{a}$ to denote elements $(a_k)_{k\in\mbN}$ of $\mcA^\mbN$. Endowing the set $\mcA$ with the discrete topology (which is clearly compact), the set $\mcA^\mbN$ becomes a compact topological space in the product topology, by Tychonoff's theorem. 
In fact, $\mcA^\mbN$ is a separable compact metric space in the product topology; see Appendix \ref{App:Measurability considerations}. 
We denote the corresponding Borel $\sigma$-algebra on $\mcA^\mbN$ by $\Sigma$. Recall that the Borel $\sigma$-algebra $\Sigma$ is the smallest $\sigma$-algebra on $\mcA^\mbN$ for which the maps $\pi_n: \mcA^\mbN  \to \mcA^n$ defined by
\begin{equation}
        \pi_n\left ((a_k)_{k\in\mbN} \right ) \ = \  (a_1, \dots, a_n)
\end{equation}
are measurable, where $\mcA^n$ has the discrete $\sigma$-algebra $2^{\mcA^n}$.  We let $\Sigma_n= \pi_n^{-1}(2^{\mcA^n}) \subset \Sigma$, and, by a slight abuse of notation, will sometimes write $\Sigma_n$ also for the power set $2^{\mcA^n}$. Since $2^{\mcA^n}$ is generated by the one point sets $\{(a_1,\dots,a_n)\}$, it follows that $\Sigma$ is generated by the \emph{cylinder sets} $\pi_n^{-1}(a_1, \dots, a_n)$, a fact that we will use freely below    
\subsection{Disordered quantum trajectories}\label{Sec2:Disordered quantum trajectories}
We shall now introduce the main object of study, namely, the disordered quantum trajectory. For any $n\in\mbN$ and $\bar{a}\in\mcA^\mbN$, let $\Vabarn:\Omega\to\matrices$ denote the random variable defined by
\begin{equation}\label{Definition of V (n) sub bar a omega}
    V^{(n)}_{\bar{a}; \omega} \ := \  V_{a_n; \theta^n(\omega)}\cdots V_{a_1; \theta(\omega)} \ .
\end{equation}
It follows that
\begin{equation}
    \sum_{\bar{a}\in\mcA^n}V_{\bar{a}}^{(n)\dagger}V_{\bar{a}}^{(n)}
    \ = \ 
    I \qquad \text{ for all $n\in \mbN$}
\end{equation}
almost surely. Note that, when $\mcA^\mbN\times \Omega$ is given the product $\sigma$-algebra $\Sigma\otimes\mcF$, the map $(\bar{a}, \omega)\mapsto V^{(n)}_{\bar{a}; \omega}$ is measurable. We shall sometimes simply write $V_{\bar{a}}^{(n)}$ to denote the random variable $\omega\mapsto V^{(n)}_{\bar{a}; \omega}$. 
For a matrix $V\in\matrices$ and $\varrho\in\mbS_d$, we define 
\begin{equation}\label{dotnotation}
    V\cdot\varrho
    \ := \ 
    \begin{cases}
        \cfrac{V\varrho V^\dagger}{\operatorname{tr}(V\varrho V^\dagger)}&\text{if } \tr{V\rho V^\dagger} \neq 0\\
        0&\text{else.}
    \end{cases}
\end{equation}
For any state $\varrho\in\mbS_d$, $\omega\in\Omega$, and $n\in\mbN$, we define $\rho^{(n)}_{\varrho;\omega}:\mcA^\mbN\to \mbS_d\cup\{0\}$ by
\begin{equation}\label{def:traj}
    \rho^{(n)}_{\varrho;\omega}(\bar{a}) \ = \  V^{(n)}_{\bar{a};\omega}\cdot \varrho \ .
\end{equation}
Note that for each fixed initial state $\varrho$, the map $(\omega,\bar{a}) \mapsto \rho^{(n)}_{\varrho;\omega}(\bar{a})$ is $(\mcF\otimes\Sigma_n)$-measurable. It will also be useful later to note that 
\begin{equation}\label{Measurability of Gamma map}
    (\varrho, \bar{a},\omega) \ \mapsto \ \rho^{(n)}_{\varrho;\omega}(\bar{a})
\end{equation}
is jointly measurable as a map $\mbS_d\times\mcA^\mbN\times\Omega \to\mbS_d\cup\{0\}$; for completeness we prove this fact in Lemma \ref{Gamma is measurable} below. When we write $\rho^{(n)}$ undecorrated, it is understood to reference the map in \eqref{Measurability of Gamma map}.
\begin{definition}\label{Definition of disordered quantum trajectory}
Let $\mcV$ be a disordered perfect Kraus measurement associated to $\mcA$. We call the sequence $\bigl(\rho^{(n)}\bigr)_{n\in\mbN}$ defined by \eqref{Measurability of Gamma map} the \emph{disordered quantum trajectory} associated to $\mcV$. When both an initial state $\varrho$ and disorder $\omega$ are fixed, we call the sequence $\bigl(\rho^{(n)}_{\varrho;\omega}\bigr)_{n\in\mbN}$ defined by \eqref{def:traj} the \emph{disordered quantum trajectory with initial state $\varrho$ and disorder $\omega$} associated to $\mcV$.
\end{definition}
Associated to a disordered quantum trajectory $\bigl(\rho^{(n)}\bigr)_{n\in\mbN}$, there is some additional data. Note that, since $\mcV\in\scrV_\mcA$ almost surely, the map $\Phi_\mcV$ defined in \eqref{Trace preservation for perfect Kraus measurement} is trace-preserving almost surely. In particular, for every $\varrho\in\mbS_d$ it holds that 
\begin{equation}
    \sum_{a\in\mcA}
    \tr{V_a \varrho V_a^\dagger}
        \ = \  
    1
\end{equation}
almost surely. Since $\operatorname{tr}(V_a\varrho V_a^\dagger)\geq 0$, it follows that
$\left(\operatorname{tr}(V_{a}\varrho V_{a}^\dagger)\right)_{a\in\mcA}$ is almost surely a probability distribution on $\mcA$. More generally, for any $\varrho\in\mbS_d$ and $n\in\mbN$, 
\begin{equation}\label{Definition of n level quantum probability}
    \mbQ^{(n)}_{\varrho}(A)
    \ := \ 
    \sum_{\bar{a}\in A} 
    \tr{
    V^{(n)}_{\bar{a}}
                \varrho
            V^{(n)\dagger}_{\bar{a}}
        }
\end{equation}
defines a random probability measure on the finite set $\mcA^n$. Furthermore, for all $n\in\mbN$ and $A\in\Sigma_n$, since $\mcV\in\scrV_\mcA$ almost surely, it follows from the cyclicity of trace and \eqref{Trace preservation of Kraus measurement} that
\begin{equation}
    \mbQ_{\varrho}^{(n+1)}(\mcA\times A)
        \ = \ 
        \mbQ_{\varrho}^{(n)}(A) \quad \text{a.s.}
\end{equation}
for any $n\in\mbN$ and $A\in\Sigma_n$.
Thus, the sequence $\bigl(\mbQ_{\varrho}^{(n)}\bigr)_{n\in\mbN}$ of random probability measures almost surely satisfies the consistency conditions required in the Daniell-Kolmogorov extension theorem. Therefore, for almost every $\omega\in\Omega$ there is a probability measure $\mbQ_{\varrho; \omega}$ on $\mcA^\mbN$ such that 
\begin{equation}\label{Quantum probability: Defining property}
        \mbQ_{\varrho;\omega}\left(
            \pi_n^{-1}(\bar{a})
        \right)
        \ = \ 
        \tr{\Vabarnomega\varrho\Vabarnomegadagger}
\end{equation}
for all $n\in\mbN$ and $\bar{a}\in\mcA^n$.\footnote{Note that $\mbQ_{\varrho;\omega}$ may fail to be a probability measure for $\omega$ in a set of measure zero at which \eqref{krausmeas} fails. This leads to no problems since the exceptional set has measure zero.} We have thus defined a map
\begin{align}\label{Random quantum probability}
    \begin{split}
        \mbQ_{\varrho}: \Omega &\to \mcP\left(\mcA^\mbN\right)\\
        \omega&\ \mapsto \  \mbQ_{\varrho;\omega} \ ,
    \end{split}
\end{align}
where $\mcP\left(\mcA^\mbN\right)$ denotes the set of Radon probability measures on $\mcA^\mbN$. Endowing $\mcP\left(\mcA^\mbN\right)$ with the Borel $\sigma$-algebra arising from the Prokhorov metric, one can show that \eqref{Random quantum probability} defines a measurable map; see Lemma \ref{The quantum probability is jointly measurable}. In particular, we may justifiably refer to $\mbQ_\varrho$ as a random probability measure. We denote by $\avg_{\varrho;\omega}\left[\cdot \right]$ the integration with respect to the measure $\mbQ_{\varrho;\omega}$. 
For a fixed initial state $\varrho\in\mbS_d$, we refer to the measure $\mbQ_{\varrho;\omega}$ as the \textit{quantum probability} associated to the tuple $(\varrho, \omega)$, and $\avg_{\varrho;\omega}$ as the associated quantum expectation. For $a_1, \dots, a_n\in\mcA$ and $\omega\in\Omega$, the quantity $\mbQ_{\varrho;\omega}\left(\pi_n^{-1}\{(a_1, \dots, a_n)\}\right)$ may be interpreted as the probability that we observe the sequence of outcomes $(a_1,\ldots a_n)$ after $n$-many measurements, given that our system is initially in state $\varrho$ and the disordered environment is configured according to $\omega\in\Omega$. 
\subsection{Asymptotic purification}\label{Sec2:Purification}
A quantum trajectory started at an initial state $\varrho$ under disorder realization $\omega$ is said to \textit{asymptotically purify} along a sequence of observed outcomes $\bar{a}\in \mcA^\mbN$ if the sequence $\bigl(\rho^{(n)}_{\varrho;\omega}(\bar{a})\bigr)_{n\in\mbN}$ asymptotically approaches the set of pure states. Equivalently, the quantum trajectory asymptotically purifies along $\bar{a}$ if the spectrum $\sigma(\rho^{(n)}_{\varrho;\omega}(\bar{a}))$ approaches the set $\{0,1\}$ as $n\to \infty$, which happens if and only if 
\begin{equation}
    \forall m\in\mbN\quad \lim_{n\to\infty} \tr{\rho^{(n)}_{\varrho;\omega}(\bar{a})}^m \ = \  1 \ .
\end{equation}
Our goal is to understand the extent to which asymptotic purification of the disordered quantum trajectory $\left(\rho^{(n)}\right)_{n\in\mbN}$ depends on the initial state, disorder realization, and the sequence of observed measurement outcomes. Towards this end, we introduce the following definition.
\begin{definition}[Purification at a particular disorder realization]\label{disordered purification}
    Given an initial state $\varrho\in\mbS_d$ and a disorder realization $\omega\in\Omega$, we say that the disordered quantum trajectory $\bigl(\rho^{(n)}_{\varrho;\omega}\bigr)_{n\in\mbN}$ \emph{asymptotically purifies at $\omega$} if
    \begin{equation}\label{Definition of purification at a particular disorder realization}
    \lim_{n \to \infty}
            \tr{\left(
                \rho^{(n)}_{\varrho; \omega}(\bar{a})
            \right)^m} \ = \  1
    \end{equation} 
    for all $m\in\mbN$ and for 
    $\mbQ_{\varrho;\omega}$-almost every $\bar{a}\in\mcA^\mbN$. 
\end{definition}
Now consider the set $X\subset\Omega$ of disorder realizations for which the quantum trajectory started at any initial state $\varrho$ asymptotically purifies with $\mbQ_{\varrho;\omega}$-probability $1$:
\begin{equation}\label{Purification event, basic version}
    X
    \ = \ 
     \bigcap_{\varrho\in\mbS_d}\bigcap_{m\in\mbN}
     \left\{
     \omega\in\Omega
        \,\,:\,\,
        \mbQ_{\varrho;\omega}\left\{\bar{a}\in\mcA^\mbN : 
        \lim_{n}
            \tr{\left(
                \rho^{(n)}_{\varrho;\omega}(\bar{a})
            \right)^m} = 1\right\}
            =
            1
    \right\} \ .
\end{equation}
Below, we prove that $X$ is a measurable set by demonstrating that purification depends continuously on the variable $\varrho\in\mbS_d$, which will allow us to replace the uncountable intersection over $\mbS_d$ in \eqref{Purification event, basic version} by the intersection over a countable dense subset of states, and therby write $X$ as a countable intersection of events in $\mcF$. In particular, we may rightfully refer to $X$ as the \textit{purification event}. Our first theorem states that, as a consequence of the ergodic structure of $(\Omega, \Pr, \theta)$, we have $\Pr(X) \in \{0,1\}$:
\begin{restatable}[]{thm}{firstresult}
\label{0-1 law for purification of disordered quantum trajectories}
Let $\mcA$ be a finite set of outcomes, let $\mcV:=\{V_a\}_{a\in\mcA}$ be any disordered perfect Kraus measurement ensemble associated to $\mcA$, and let $X_\mcV$ denote the corresponding purification set of disordered realizations for $\mcV$. Then $X_{\mcV}\in\mcF$ and $\Pr(X_\mcV)\in\{0, 1\}$.
\end{restatable}
{\raggedright We} say that a disordered perfect Kraus measurement ensemble $\mcV$ associated to $\mcA$ \emph{ asymptotically purifies} if $\Pr(X_\mcV) = 1$. The above theorem establishes that the asymptotic purification of the disordered quantum trajectory depends only on the function $\mcV:\Omega\to\scrV_\mcA$. We are therefore led to ask what properties of $\mcV$ determine whether it asymptotically purifies. 
A first step in this direction is recalling the main theorem of Maassen and K\"ummerer in the non-disordered setting \cite{maassen2006purification}, which states that a non-disordered quantum trajectory started at any initial state asymptotically purifies along quantum almost every sequence $\bar{a}$ of observed outcomes if and only if the Kraus measurement operators $\{V_a\}_{a\in\mcA}$ satisfy the following property. 
\begin{description}
\hypertarget{pur}{}
\item[\pur] 
There is $n\in\mbN$ such that any projection $p$ that satisfies
$pV_{a_1}^\dagger\ldots V_{a_n}^\dagger V_{a_n}\ldots V_{a_1}p \propto p$ for all $a_1,\ldots a_n \in \mcA$ is necessarily rank 1. 
\end{description}
The notation $A\propto B$ indicates that $A$ is proportional to $B$, i.e., there is $\lambda\in\mbC$ such that $A = \lambda B$. Our next theorem establishes an analogous equivalent condition for the present disordered situation. 
\begin{restatable}[]{thm}{secondresult}
\label{Main theorem}
Let $\mcA$ be a finite set of observable outcomes and let $\mcV$ be a disordered perfect Kraus measurement associated to $\mcA$. Then $\mcV$ asymptotically purifies if and only if the following condition holds:
\begin{description}
\hypertarget{dpur}{}
\item[\dpur] There is $n\in \mbN$ and an event $S\subset \Omega$ of positive $\Pr$-probability such that for each $\omega \in S$, any projection $p$ satisfying $pV_{\bar{a}; \omega}^{(n)\dagger} V_{\bar{a}; \omega}^{(n)} p \propto p$ for all $\bar{a}\in\mcA^n$ is necessarily rank 1.
\end{description}
\end{restatable}
\begin{remark}\label{rem:constructive} This result provides a constructive criterion for asymptotic purification: to show that $\mcV$ asymptotically purifies, it suffices to demonstrate the existence of a subset of $\Omega$ with positive (but possibly very small) probability satisfying the condition in \hyperlink{dpur}{(d-Pur)}. If this holds, it can in principle be verified by a finite computation. See Remark \ref{rem:technicalconstructive} below for a more precise statement of this criterion.
\end{remark}
In the non-disordered setting of \cite{maassen2006purification}, a projection $p$ satisfying
\begin{equation}\label{Deterministic dark subspaces}
    \text{for all $n\in\mbN$ and $a_1, \dots, a_n\in\mcA$,} \quad
    pV_{a_1}^\dagger\cdots V_{a_n}^\dagger V_{a_n}\cdots V_{a_1} p \ \propto \ p
\end{equation}
is called a \textit{dark projection}, and its range is called a \textit{dark subspace}. Thus, \pur says that any dark projection is necessarily rank 1, and the main theorem of \cite{maassen2006purification} may be understood as saying that asymptotic purification occurs if and only if all dark projections are of rank 1. Following this terminology, we make the following definition.
\begin{definition}[Dark projections and subspaces]
    Let $\mcV$ be a disordered perfect Kraus measurement associated to $\mcA$. Given a disorder realization $\omega\in\Omega$, a projection $p$ of rank $r\geq 2$ is called a \emph{dark projection for $\mcV_\omega$} if 
    \begin{equation}\label{Disordered dark subspaces}
    \text{for all $n\in\mbN$ and $\bar{a}\in\mcA^n$,} \quad
    p V_{\bar{a}; \omega}^{(n)\dagger}V_{\bar{a}; \omega}^{(n)} p
    \ \propto \ p \ ,
    \end{equation}
    and the range of such $p$ is called a \emph{dark subspace}. The set of all dark projections of rank $r\geq 2$ for $\mcV_\omega$ is denoted $\scrD_\omega$.
\end{definition}
Note that in the above definition we exclude projections $p$ with rank $r = 1$, since for all such projections \eqref{Disordered dark subspaces} holds trivially. Note also that we adopt the same terminology as in \cite{maassen2006purification}, i.e., we call a projection $p$ satisfying \eqref{Disordered dark subspaces} a dark projection. We note that the notion of a dark projection as introduced in \cite{maassen2006purification} may be understood as a dark projection according to our definition where $|\Omega| = 1$ and $\theta$ is the identity map. 
We write $\scrD$ to denote the set-valued map $\omega\mapsto \scrD_\omega$ and, for integers $r\geq 2$, we let $\scrD^{(r)}$ denote the set-valued map $\omega\mapsto \scrD_\omega\cap\scrP_r$, so $\scrD^{(r)}$ is the subset of $\scrD$ consisting of rank $r$ dark projections. One can show that $\Pr$-almost surely $\scrD_\omega^{(r)}$ is a nonempty compact set within $\matrices$ and that, when the collection of compact subsets of $\matrices$ is given the Borel $\sigma$-algebra arising from the Hausdorff metric, the association $\omega\mapsto \scrD_\omega^{(r)}$ is measurable. In particular, $\scrD^{(r)}$ defines a compact set-valued random variable. With this in hand, we may state the following refinement of Theorem \ref{Main theorem}.
\begin{restatable}[]{thm}{thirdresult}
\label{Main theorem'}
    Let $\mcA$ be a finite set of observable outcomes, let $\mcV$ be a disordered Kraus measurement associated with $\mcA$. Then the map $\omega \mapsto \scrD_\omega$ is $\mcF$-measurable, and, up to a set of measure zero, the event that there are no dark projections is equal to the purification event $X_\mcV$, i.e., $ \left\{\scrD = \emptyset\right\} = X_\mcV$ almost surely. In particular, $\mcV$ asymptotically purifies if and only if $\Pr\left(\scrD = \emptyset\right) > 0$.
\end{restatable}
\begin{remark}\label{Rmk:Ergodic decomposition for stationary}
If $\Omega$ is a complete separable metric space, in particular if $\Omega$ is a Lebesgue space, then we may extend the above results to apply to stationary but not necessarily ergodic dynamics by means of the Ergodic Decomposition Theorem (see \cite[Theorem 5.1.3]{Viana_Oliveira_2016}.
Indeed, if we assume only that $\mbP\circ\theta = \mbP$,  then this theorem 
asserts that there is an almost sure partition $\mcD$ of $\Omega$ into $\theta$-invariant measurable subsets $D\in\mcD$, 
\begin{equation}
    \Omega 
    \ = \ 
    \bigsqcup_{D\in\mcD}
    D\ ,
\end{equation}
a probability measure $\hat{\mbP}$ on $\mcD$, and for each $D\in \mcD$ a probability measures $\mu_D$ on $D$ which is ergodic and invariant for $\theta$, such that, for any event $E$, it holds that 
\begin{equation}
    \mbP(E) \ = \  \int_\mcD \mu_D(E)\,d\hat{\mbP}(D) \ .
\end{equation}
Thus, because the dynamics of the trajectory is determined only by the particular disorder realization $\omega\in\Omega$, and hence $\omega\in D$ for some $D\in\mcD$, the above implies that, upon choice of $\omega\in\Omega$, we may replace our dynamics $\big(\Pr, \theta:\Omega\to\Omega\big)$ by the ergodic dynamical system $\big(\mu_D, \theta:D\to D\big)$. 

That is, the dynamics may be conceived of as being specified in two steps: first, a random choice of component $D\in \mcD$ according to the distribution $\hat{\mbP}$, followed by a choice of $\omega\in D$ (and thus the full dynamics $n\mapsto \theta^n(\omega) \in D$) by the ergodic measure $\mu_D$.  In particular, each of the above results describes the behavior of a quantum trajectory under a sequence of stationary random generalized measurements, after conditioning on the choice of ergodic component.
\end{remark}
\subsection{Examples} 
We now present some examples to illustrate disordered perfect Kraus measurements both with dark subspaces and without. 
\begin{example}\label{Example:Trivially purifying example}
 Let $\mbU_3$ be the group of $3\times 3$ unitary matrices with respect to the standard basis on $\mbC^3$, and let $\nu$ denote the normalized Haar measure on $\mbU_3$. Consider the probability space
\begin{equation}
    \scrU_3 \ = \  \prod_{i\in\mbZ}\mbU_3 \ ,
\end{equation}
with the product measure $\mu = \bigotimes_{i\in \mbZ} \nu$.  For $\omega\in\scrU_3$, we write $\omega_i$ for the $i$-the entry of $\omega$, and let $U_i:\scrU_3 \to \mbU_3$ denote the map $U_i(\omega)=\omega_i$. We note that under $\mu$ the random matrices $\{U_i\}_{i\in \mbZ}$ are independent and identically distributed according to Haar measure $\nu$.  Let $\theta:\scrU_3\to\scrU_3$ be the left shift,
\begin{align}
        (\omega_i)_{i\in\mbZ}\xmapsto{\theta} (\omega_{i+1})_{i\in\mbZ} \ .
\end{align}
It is a standard fact that $\theta$ is strongly mixing, hence ergodic, for $\mu$. For $j= 1, 2, 3$, let $\ket{\omega_i, j}$ denote the $j$th column vector of $\omega_i$. Let $\mcA = \{a, b\}$ and consider the disordered perfect Kraus measurement on $\mcV$ on $\mcA$ defined by 
    \begin{align}
        \begin{split}
             \mcV:&\scrU_3 \to \scrV_\mcA , \quad \omega \xmapsto{\mcV} 
        \Big\{
            V_{a; \omega}, \,
            V_{b; \omega} 
        \Big\}\ , \\
        & V_{a; \omega} \ = \  \ket{\omega_{0}, 1}\bra{\omega_0, 1} + \ket{\omega_{0}, 2}\bra{\omega_{0}, 2}\\
        & V_{b; \omega} \ = \  \ket{\omega_0, 3}\bra{\omega_0, 3} \ .
        \end{split}
    \end{align}
We claim that $\mcV$ asymptotically purifies. Indeed $\mcV$ satisfies the stronger property of \emph{eventual purification}, i.e., for almost every $\omega$, with $\mathbb{Q}_{\varrho;\omega}$ probability one, there is $n_0\in \mbN$ such that $\rho^{(n)}$ is a pure state for all $n\ge n_0$.  Indeed, with probability one  the outcome $b$ is eventually realized in one of the measurements, after which time the state $\rho^{(n)}$ is rank one since the Kraus map $V_{b;\omega}$ is rank one. 
To verify that the outcome $b$ eventually occurs, let $\tau_b:\mcA^\mbN\to\mbN\cup\{\infty\}$ be the random variable defined by 
\begin{equation}
        \tau_b(\bar{a})
        \ := \ 
        \inf\{n\in\mbN\,\,:\,\, a_n = b\} \ ,
\end{equation}
where $a_n$ denotes the $n$th coordinate of $\bar{a} \in \mcA^\mbN$ and we take $\inf\emptyset = \infty$. We claim that for any state $\varrho\in \mbS_3$ and $\mu$-almost every $\omega\in\Omega$, we have that 
\begin{equation}\label{Eqn:Trivially purifying example, eqn 1}
    \mbQ_{\varrho; \omega}\Big[
        \tau_b < \infty
    \Big]
    \ = \     1 \ .
\end{equation}
Assuming this to be the case, it follows immediately that $\mcV$ eventually purifies: on the event that $\tau_b < \infty$, we have that $\rho^{(n)}_{\varrho; \omega}$ is a pure state for $n\ge \tau_b$. 

To prove that \eqref{Eqn:Trivially purifying example, eqn 1} holds, we show $\mbQ_{\varrho; \omega}\Big[ \tau_b = \infty
\Big] = 0$ for $\mu$-almost every $\omega$. 
Since $\mbQ$ takes values in $[0, 1]$, it suffices to show that 
        \begin{equation}
            \int_{\scrU_3}
            \mbQ_{\varrho; \omega}\Big[
            \tau_b = \infty
        \Big]
        \ d\mu(\omega)
        \ = \ 
        0.
        \end{equation}
We note further that $\tau_b(\bar{a}) = \infty$ if and only if $\bar{a}$ is a sequence consisting of all $a$'s.  Thus, by dominated convergence, it suffices to show that 
\begin{equation}\label{Eqn:Trivially purifying example, eqn 3}
         \lim_{n\to\infty}  \int_{\scrU_3}\tr{R_{n;\omega}} \,d\mu(\omega) \ = \  0 \ ,
\end{equation}
where $ R_{n;\omega} = V_{a;\theta^n(\omega)} \cdots V_{a;\theta(\omega)} \varrho V_{a;\theta(\omega)}\cdots V_{a;\theta^n(\omega)}$.
    Note that $V_{a;\theta^i(\omega)} = U_i(\omega) P U_i^\dagger(\omega)$ with $P$ the rank-two projection onto the first two elementary basis vectors.  Since $U_n$ is unitary, we have
    \begin{equation}
        \tr{R_n} \ = \  \tr{P U_n^\dagger R_{n-1} U_n P} \ .
    \end{equation}
    Because $R_{n-1}$ depends only on $U_1,\ldots,U_{n-1}$, and is thus independent of $U_n$, it follows that
    \begin{equation}\label{Rn integral}
    \int_{\scrU_3} \tr{R_{n;\omega}} d \mu(\omega) \ = \ \int_{\scrU_3} \int_{\mbU_3} \tr{PU^\dagger R_{n-1;\omega} U P} d\nu(U) d \mu(\omega) \ = \ \frac{2}{3} \int_{\scrU_3}\tr{ R_{n-1;\omega}}  d \mu(\omega) \ ,
    \end{equation}
    where in the last step we have used the identity\footnote{To see \eqref{Haar identity}, note that the expression is linear in $A$ and invariant under unitary conjugation (by left invariance of Haar measure).  This fixes the result to be proportional to $\tr{A}$; computing the integral for $A=I$ yields the constant of proportionality.}
    \begin{equation}\label{Haar identity}
    \int_{\mbU_3} \tr{P U^\dagger A U P} d\nu(U) \ = \ \frac{2}{3} \tr{A} \ .
    \end{equation}
    Iterating \eqref{Rn integral}, we find that $\int_{\scrU_3}\tr{R_{n}} \,d\mu = \left (\frac{2}{3} \right )^n$, and thus that \eqref{Eqn:Trivially purifying example, eqn 3} holds and eventual purification holds for $\mcV$.
\end{example}
\begin{example}\label{Example:Random dark subspaces example}
    Let $\scrU_3$, $\theta$, and $\mcA$ be as in the previous example, but instead consider the slightly different disordered perfect Kraus measurement ensemble $\mcW$ given by 
    \begin{align}
        \begin{split}
             \mcW:\scrU_3 &\to \scrV_\mcA , \quad  \omega \xmapsto{\mcW}
        \Big\{
            W_{a; \omega}, \,
            W_{b; \omega} 
        \Big\}\ , \\
        & W_{a; \omega} \ = \  \ket{\omega_{1}, 1}\bra{\omega_0, 1} + \ket{\omega_{1}, 2}\bra{\omega_{0}, 2}\  , \\
        & W_{b; \omega} \ = \  \ket{\omega_1, 3}\bra{\omega_0, 3} \ .
        \end{split}
    \end{align}
We note that 
    \begin{equation}
        W_{a; \theta(\omega)}W_{b; \omega} \ = \  W_{b; \theta(\omega)}W_{a; \omega} \ = \  0
    \end{equation}
    for every $\omega\in\scrU_3$. If we let $p$ be the random projection defined by $p_\omega = \ket{\omega_1, 1}\bra{\omega_1, 1} + \ket{\omega_1, 2}\bra{\omega_1, 2}$, we see that $p$ is a projection of rank 2 satisfying
    \begin{align}
        p_{\omega}W_{\bar{a}; \omega}^{(n), \dagger} W_{\bar{a}; \omega}^{(n)}p_\omega
        \ &\propto \ p_\omega
    \end{align}
    for all $n\in\mbN$, $\bar{a}\in\mcA^\mbN$, and $\omega\in\scrU_3$. That is, $p_\omega\in \scrD_\omega$ for all $\omega\in\scrU_3$. Furthermore, $p_\omega$ is non-deterministic and highly dependent on the disorder. Note that in this example, the random ensembles $\mcW$ and $\mcW_{\theta(\cdot)}$ are identically distributed \emph{but not independent}. 
\end{example}
\begin{example}\label{Example:Harder to prove purifying example}
    We modify Example \ref{Example:Trivially purifying example} to obtain an example in which asymptotic purification holds without eventual purification. Let $\mbU_4$ be the group of $4\times 4$ unitary matrices with respect to the standard basis on $\mbC^4$, and, as before, consider 
    \begin{equation}
        \scrU_4 
        \ = \ 
        \prod_{i\in\mbZ}\mbU_4
    \end{equation}
    with the product measure $\mu$ arising from the normalized Haar measure on $\mbU_4$. As before, we let $\theta:\scrU_4\to\scrU_4$ be the left shift, defined as in Example \ref{Example:Trivially purifying example}. We consider now the disordered perfect Kraus measurement $\mcV$ on $\mcA = \{a, b\}$ defined by 
    \begin{align}
        \begin{split}
            \mcV:&\scrU_4\to\scrU_4\ , \quad \omega\xmapsto{\mcV} \{V_{a; \omega}, V_{b; \omega}\} \ , \\
            & V_{a; \omega} \ = \  \ket{\omega_{0}, 1}\bra{\omega_0, 1} + \ket{\omega_{0}, 2}\bra{\omega_{0}, 2} \ , \\
            & V_{b; \omega} \ = \  \ket{\omega_0, 3}\bra{\omega_0, 3} + \ket{\omega_0, 4}\bra{\omega_0, 4} \ ,
        \end{split}
    \end{align}
    where $\ket{\omega_0,j}$, $j=1,2,3,4$, are the columns of $\omega_0$. The measurement operators associated with each outcome are both rank two projections.  Nonetheless, we claim that $\mcV$ purifies.  As above, we will use the notation 
    \begin{equation}
        V_{a;j} \ = \  V_{a;\theta^j(\omega)} \quad \text{and} \quad V_{b;j} = V_{b;\theta^{j}(\omega)} \ .
    \end{equation}
    
By Theorem \ref{Main theorem'}, to prove that $\mcV$ purifies, it suffices to prove, for almost every $\omega$, that any projection $p$ that satisfies  \eqref{Disordered dark subspaces} is necessarily rank one.  Fix $\omega$ and let $p$ be a projection that satisfies \eqref{Disordered dark subspaces} \textemdash \ so $p$ would be a dark projection if $\tr{p}\ge 2$. We have, in particular, the set of six equations 
    \begin{equation}\label{Eqn:Harder to prove purifying example, eqn 1}
    \begin{rcases*}
        \lambda_{c} p \\
        \lambda_{\bar{c}}p 
    \end{rcases*}
    \ = \  \begin{cases}
        p V_{c;1}p \ , & c \in \mcA \ ,\\
        pV_{c_1;1} V_{c_2;2} V_{c_1;1} p \ , & \bar{c}\in \mcA^2 \ ,
    \end{cases}
    \end{equation}
    where the various numbers $\lambda_{ \sharp}$, $\sharp \in \mcA\cup \mcA^2$, are in $[0, 1]$ and the simplified form of the above equations follows from the fact that each of the Kraus operators in $\mcV$ is a projection. Recall that we have fixed $\omega$ and are considering these objects only for that particular value of the disorder. Both $p$ and the numbers $\lambda_{\sharp}$ depend on the disorder point $\omega$.  They could, in fact, be chosen to be measurable functions of $\omega$, but that is not important for our argument. 

    Because $V_a + V_b=I$, the equations \eqref{Eqn:Harder to prove purifying example, eqn 1} are not independent. Indeed, we have 
    \begin{equation}
        \lambda_{a}+\lambda_b\ = \ 1 \ , \quad \lambda_{a,a}+ \lambda_{a,b} \ = \  \lambda_a \ , \quad \text{and} \quad \lambda_{b,a} + \lambda_{b,b}\ = \ \lambda_b \ . 
    \end{equation}
    From the first of these three equations, we see that at least one of $\lambda_a$ or $\lambda_b$ must be non-zero.  For definiteness, we take
    \begin{equation}
        c
        \ := \ 
        \begin{cases}
            a & \text{if }\lambda_{a}>0 \ ,\\
            b &\text{if }\lambda_{a} = 0 \ ,
        \end{cases}
    \end{equation}
    and note that in either case $\lambda_c >0$. Furthermore, for any $\varphi\in \operatorname{ran}\left (p \right )$, we have that
        \begin{equation}\label{Eqn:Harder to prove purifying example, eqn 2}
        \|\varphi\|
        \ = \ 
        \sqrt{\lambda_{c}}\cdot\|V_{c;1}\varphi\| 
        \quad\text{and}\quad 
         \|\varphi\|
        \ = \ 
        \sqrt{\lambda_{d, c}}\cdot \|V_{d;2} V_{c;1}\varphi\| \ , \quad d\in \mcA \ .
    \end{equation}

    Since $V_{a;1}$ and $V_{b;2}$ are both rank two projections, it follows from \eqref{Eqn:Harder to prove purifying example, eqn 1} that $\operatorname{rank}(p)\le 2$. Suppose that $\operatorname{rank}(p)= 2$. Since $\operatorname{rank}(p) = \operatorname{rank}(V_{c;1})=2$, it follows from the first of the two equations \eqref{Eqn:Harder to prove purifying example, eqn 2} implies that 
    \begin{align}
        \begin{split}
            V_{c;1}\vert_{\operatorname{ran}\left( p \right)}:\operatorname{ran}\left(p\right)\xrightarrow[]{}\operatorname{ran} \left(V_{c;1}\right)
        \end{split}
    \end{align}
    is bijective. In particular, for any $\psi\in\operatorname{ran}\left(V_{c;1}\right)$, there is a unique $\varphi\in\operatorname{ran}(p)$ with $\psi = V_{c;1}\varphi$. Then, from \eqref{Eqn:Harder to prove purifying example, eqn 2} we find that
    \begin{equation}
        \begin{rcases*}
            \|\psi\|\\
            \|V_{d;2}\psi\|
        \end{rcases*}
        \ = \ 
        \begin{cases}
            \sqrt{\lambda_{c}}\cdot \|\varphi\|\\
            \sqrt{\lambda_{d,c}}\cdot \|\varphi\| \ , & d \in \{a,b\}
        \end{cases}
    \end{equation}
    whence we conclude that
    \begin{equation}\label{Eqn:Harder to prove purifying example, eqn 3}
        \|V_{d;2}\psi\|
        \ = \ 
        \sqrt{\frac{\lambda_{d,c}}{\lambda_{c}}}\cdot \|\psi\| \ , \quad d \in \{a,b\} \ .
    \end{equation}
    In particular, the two linear maps
    \begin{align}\label{Eqn:Harder to prove purifying example, eqn 4}
        \begin{split}
            V_{d;2}\vert_{\operatorname{ran}\left( V_{c;1} \right)}:
            \operatorname{ran}\left( V_{c;1} \right)
            \xrightarrow[]{}
            \operatorname{ran} \left(V_{d;2}\right) \ , \quad d\in \{a,b\}
        \end{split}
    \end{align}
    are each proportional to isometries. 

    We have shown that the event that $\mcV$ has a dark projection is contained in the event     $$ 
    \bigcup_{\bar{c}\in \mcA^2} E_{\bar{c}} \, \quad \text{where} \quad E_{\bar{c}} = \{ V_{c_2;2}\vert_{\operatorname{ran}(V_{c_1;1})}:\operatorname{ran}\left ( V_{c_1;1} \right ) \to \operatorname{ran} \left (V_{c_2;2} \right ) \text{ is proportional to an isometry}\} \ .$$
    We note that for each $\bar{c}\in \mcA^2$, the two operators $V_{c_2,2}$ and $V_{c_1,1}$ are independent and each distributed according to $V_{a,1}$.  Thus $\mu(E_{\bar{c}})=\mu(E_{(a,a)})$ for each $\bar{c}\in \mcA^2$. Thus it suffices to prove that $\mu(E_{(a,a)}=0$.

    Let 
     \begin{equation}
        q \ = \  \left(
            \begin{array}{cccc}
                1 & 0 & 0 & 0\\
                0 &  1 & 0 & 0\\
                0 & 0 &0 &0 \\
                0 & 0& 0& 0
            \end{array}
        \right),
    \end{equation} 
    and note that $V_{a,j}= U_j q U_j^\dagger$ for $j=1,2$. Thus, since $U_1$ and $U_2$ are unitary, $\operatorname{ran} (V_{a,1}= U_1 \operatorname{ran}(q)$ and
    $$ E_{(a,a)} \ = \ \{ q U_1\vert_{\operatorname{ran}(q)}:\operatorname{ran}(q) \to \operatorname{ran}(q) \ \text{ is proportional to an isometry} \} \ .$$
    That is $\omega \in E_{(a,a)}$ if and only if
    \begin{equation}\label{Eqn:Harder example codimension}
    q U_1(\omega)^\dagger q U_1(\omega) q \ \propto \ q  \ .
    \end{equation}
    The set of unitary matrices satisfying \eqref{Eqn:Harder example codimension} is a variety of positive codimension and thus has Haar measure zero.  Thus $\mu(E_{(a,a)})=0$.  This concludes the proof of the  asserted asymptotic purification.
    We remark that this example has an obvious generalization to the case where $\mbU_4$ is replaced by $\mbU_d$ for arbitrary $d\in\mbN$, $\mcA$ is replaced by a set with cardinality $k$ for some suitable $k\in\mbN$, and for each $a\in\mcA$, $V_{a; \omega}$ is defined to be the projection onto the span of some finite set of column vectors of $\omega_0$. By appropriately modifying the method above, one can show that the resulting disordered perfect Kraus measurement also purifies. 
\end{example}
\subsection{Organization of proofs}\label{Sec2:Organization of proofs}
In \S \ref{Sec:Purification event}, we establish that the so-called purification event $X$ defined in \eqref{Purification event, basic version} is indeed an element of the $\sigma$-algebra $\mcF$. Along the way towards this, we establish, as in \cite{maassen2006purification}, that, almost surely, the moments of the disordered quantum trajectory $\left(\rho^{(n)}\right)_{n\in\mbN}$ have the structure of a submartingale, which allows us to rewrite the definition of $X$ in a simpler form. By leveraging this simple form, we prove Corollary \ref{Purification is an event}, which states that $X\in\mcF$. Having established this necessary technical result, we are able to prove Theorem \ref{0-1 law for purification of disordered quantum trajectories}. 
In \S \ref{Sec:Dark subspaces}, we turn to the proofs of Theorems \ref{Main theorem} and \ref{Main theorem'}. Because Theorem \ref{Main theorem} is a corollary of Theorem \ref{Main theorem'}, we only prove Theorem \ref{Main theorem'}. To do this, we introduce the definition of a random compact set and establish that the set of dark projections for the disordered quantum trajectory fulfils the requirements of this definition. Along the way to that, we establish some elementary properties of the set of dark subspaces which is useful in the proof of Theorem \ref{Main theorem'}. The remainder of this section is similar in spirit to the proof of \cite[Lemma 3]{maassen2006purification}: loosely speaking, we prove that, given $\Pr(X) = 0$, for almost every $\omega\in\Omega$, we can construct a dark projection $p$ for $\mcV_\omega$. This allows us to conclude Theorem \ref{Main theorem'}.
In Appendix \ref{App:Measurability considerations}, we establish some technical results needed mostly for the arguments in \S \ref{Sec:Purification event}.

%
%

\section{The purification event}\label{Sec:Purification event}
Recall that the purification set $X$ was defined to be the subset of $\Omega$ such that the quantum trajectory $(\rho^{(n)})_{n\in\mbN}$ asymptotically purifies according to Definition \ref{disordered purification}, and was shown to be given by the expression 
\begin{equation}\label{eq:X}
        X 
            \ = \  \bigcap_{\varrho\in\mbS_d}\bigcap_{m\in\mbN}
                \left\{
                    \omega\in\Omega \ : \ 
                        \mbQ_{\varrho;\omega}   
                            \left\{
                                \bar{a}\in\mcA^\mbN : \lim_{n\to\infty} \tr{\left(\rho^{(n)}_{\varrho;\omega}(\bar{a})\right)^m} = 1
                            \right\}
                \ = \ 1
                \right\} \ .
\end{equation}
As written in \eqref{eq:X}, $X$ is expressed as an uncountable intersection, and hence its measurability is unclear.  Thus, our first order of business is to demonstrate an expression for $X$ that manifests its measurability. To do this, we start with one of the key observations in \cite{maassen2006purification}: namely, that the $m^{\textrm{th}}$ moments of a quantum trajectory, viewed as $\mcA^\mbN$-random variables, form a submartingale with respect to the filtration $(\Sigma_n)_{n\in\mbN}$. The first step towards proving the measurability of $X$ is investigating how that observation generalizes to the current disordered context.
\subsection{Moments as submartingales}\label{Sec3:Martingales}
For any $\omega\in\Omega$, let $M_{\varrho;\omega}^{(m, n)}$ denote the $\mcA^\mbN$-random variable
        \begin{align}\label{Submartingale definition}
                    \begin{split}
                        M_{\varrho;\omega}^{(m, n)}:\mcA^\mbN &\to [0, 1]\\
                    M_{\varrho;\omega}^{(m, n)}(\bar{a}) &\ = \  \operatorname{tr}\left(\left(\rho^{(n)}_{\varrho;\omega}(\bar{a})\right)^m\right)
                    \end{split}
        \end{align}
where $\varrho\in\mbS_d$ and $m, n\in\mbN$. That is, $M_{\varrho; \omega}^{(m, n)}$ is the $m$th moment of $\rho^{(n)}_{\varrho; \omega}$. As before, we sometimes write $M^{(m, n)}_{\varrho}$ or simply $M^{(m, n)}$ to denote the function of all parameters implied by \eqref{Submartingale definition}. Because $\rho^{(n)}$ is jointly measurable in each of its variables and only depends on the first $n$ observed outcomes, we have that $M^{(m, n)}$ is $\big(\mcB(\mbS_d)\otimes\Sigma_n\otimes\mcF\big)$-measurable for all $m, n\in\mbN$. The following proposition establishes the main technical result of this section.
\begin{prop}\label{Submartingale convergence}
Let $\varrho\in\mbS_d$ and $m\in\mbN$. Then, for almost every $\omega\in \Omega$, $(M_{\varrho;\omega}^{(m, n)})_{n\in\mbN}$ is a bounded submartingale with respect to the filtration $(\Sigma_n)_{n\in\mbN}$ and the probability measure $\mbQ_{\varrho; \omega}$. In particular for every initial state $\varrho$ and almost every $\omega\in\Omega$, $\lim_{n\to\infty}M_{\varrho;\omega}^{(m, n)}$ exists on a set of full $\mbQ_{\varrho;\omega}$-probability.  
            \end{prop}
   
                \begin{proof}
                We proceed as in \cite{maassen2006purification}. Fix $\varrho\in\mbS_d$. That $M^{(m, n)}_{\varrho;\omega}$ is $ \Sigma_n$ measurable is clear. To see that the submartingale property holds, we use \cite[Inequality 56]{nielsen2001characterizing}. This inequality implies that, for any $\rho\in\mbS_d$ and $\mcW\in\scrV_\mcA$, one has 
                \begin{equation}\label{eq:nielseninequality}
                    \operatorname{tr}(
                        \rho^m)
                        \ \leq \ 
                        \sum_{W\in\mcW}
                        \operatorname{tr}\left(W\rho W^\dagger\right)
                        \operatorname{tr}\big((W\cdot\rho)^m\big) \ .
                \end{equation}
                Thus for every $\bar{a}\in\mcA^\mbN$, noting that $\mcV_{\theta^{n+1}(\omega)}\in\scrV_\mcA$ almost surely, we find that
                \begin{equation}\label{eq:submartingaleprop}
                \begin{aligned}
                    M^{(m, n)}_{\varrho; \omega}(\bar{a}) \ = \  \operatorname{tr}\left(\rho^{(n)}_{\varrho; \omega}(\bar{a})^m\right)
                        &\ \leq \
                        \sum_{a\in\mcA}
                        \operatorname{tr}\left(
                                V_{a;\theta^{n+1}(\omega)}\rho^{(n)}_{\varrho; \omega}(\bar{a})V_{a;\theta^{n+1}(\omega)}^\dagger
                        \right)
                        \operatorname{tr}\left(
                                \left(
                                    V_{a;\theta^{n+1}(\omega)}\cdot\rho^{(n)}_{\varrho;\omega}(\bar{a})
                                \right)^m
                            \right)\\
                        &\ = \ 
                        \avg_{\varrho;\omega}
                            \left[
                                M^{(m, n+1)}_{\varrho; \omega}\,\Big\vert\, \Sigma_n
                            \right]
                        (\bar{a}) \ ,
                \end{aligned}
                \end{equation}
                where $\avg_{\varrho;\omega}[\ \cdot \ \vert\ \Sigma_n]$ denotes the $\mbQ_{\varrho;\omega}$-conditional expectation with respect to the $\sigma$-algebra $\Sigma_n$. Thus $(M_{\varrho;\omega}^{(m, n)})_{n\in\mbN}$ is a submartingale. It is bounded since $M^{(m, n)}_{\varrho; \omega}\in[0, 1]$ for all $n, m\in\mbN$.  By  the martingale convergence theorem (e.g., \cite[Theorem 6.18]{kallenberg1997foundations}), the limit $\lim_{n}M_{\varrho; \omega}^{(m, n)}$ exists $\mbQ_{\varrho; \omega}$-almost surely. 
                \end{proof}
By the above proposition, for every $\varrho\in\mbS_d$, $m\in\mbN$, and almost every $\omega\in\Omega$, the function
\begin{align}\label{Submartingale limit function definition}
\begin{split}
        M^{(m,\infty)}_{\varrho; \omega}: \mcA^\mbN &\to [0, 1] \\
        \bar{a} &\ \mapsto \ \lim_{n\to\infty} M_{\varrho;\omega}^{(m, n)}(\bar{a})
        \end{split}
\end{align}
is $\mbQ_{\varrho; \omega}$-almost surely defined; to make it everywhere defined, let $M^{(m,\infty)}_{\varrho; \omega}(\bar{a}) = 0$ for those $\bar{a}\in\mcA^\mbN$ such that $M^{(m,\infty)}_{\varrho; \omega}(\bar{a})$ is \textit{a priori} undefined. Note that, since $ M^{(m,n)}$ is $\big(\mcB(\mbS_d)\otimes\Sigma\otimes\mcF\big)$-measurable, the limit $M^{(m,\infty)}$ is also $\big(\mcB(\mbS_d)\otimes\Sigma\otimes\mcF\big)$-measurable.
Equipped with Proposition \ref{Submartingale convergence}, we may rewrite the purification set $X$ as
\begin{equation}\label{Purification event, basic version, for proof of Theorem 1}
    X \ = \  \bigcap_{\varrho\in\mbS_d}\bigcap_{m\in\mbN}\left\{\omega: \mbQ_{\varrho;\omega}\left(M^{(m,\infty)}_{\varrho;\omega} = 1\right) = 1
    \right\} \ .
\end{equation}
Since $M^{(m, n)}$ takes values in $[0,1]$ for all $m\in\mbN$ and $n\in\mbN\cup\{\infty\}$, we have that
\begin{equation}
 \mbQ_{\varrho;\omega}\left(
    \big\{
        \bar{a}\in\mcA^\mbN\,:\, M_{\varrho;\omega}^{(m,\infty)}(\bar{a}) = 1)
    \big\}
    \right)  =  1 \quad 
    \text{if and only if
    $\mbE_{\varrho;\omega}\left[ M_{\varrho;\omega}^{(m,\infty)}\right]=1$}   
\end{equation}
for all $\omega\in\Omega$ and $\varrho\in\mbS_d$. We may therefore rephrase the definition of $X$ as
\begin{equation}
    X \ = \  \bigcap_{\varrho\in\mbS_d}\bigcap_{m\in\mbN}\left\{\omega: \avg_{\varrho;\omega}\left[M^{(m,\infty)}_{\varrho;\omega}\right] = 1\right\} \ . 
\end{equation}
A further reduction of the description of the purification set $X$ is provided in the following lemma.
\begin{lemma}\label{Alternative definitions of purification}
The following equality holds:
    \begin{equation}\label{Alternative definitions of purification, Equation}
    X
        \ = \ 
        \bigcap_{\varrho\in\mbS_d}
    \left\{\omega\,:\,
    \mbE_{\varrho; \omega}
        \left[ 
            M_{\varrho; \omega}^{(2,\infty)}
        \right]
            =
        1
    \right\} \ .
    \end{equation}
\end{lemma}
\begin{proof}
    Fix $\varrho\in\mbS_d$, $m\in\mbN$, and $\omega\in\Omega$ and observe that $
   M_{\varrho;\omega}^{(2,\infty)}(\bar{a}) = 1 $ if and only if $M_{\varrho;\omega}^{(m,\infty)}(\bar{a}) = 1$  for all $m\in\mbN$. This is indeed the case since, if $M_{\varrho;\omega}^{(2,\infty)}(\bar{a}) = 1$, then the spectrum $\sigma\bigl(\rho^{(n)}_{\varrho;\omega}(\bar{a})\bigr)$ of $\rho^{(n)}_{\varrho;\omega}(\bar{a})$ converges to $\{0, 1\}$ with 1 being a non-degenerate eigenvalue. To see this, enumerate the eigenvalues of $\rho^{(n)}_{\varrho; \omega}$ as $0\leq \lambda_1\bigl(\rho^{(n)}_{\varrho; \omega}\bigr)\leq\dots\leq \lambda_d\bigl(\rho^{(n)}_{\varrho; \omega}\bigr)\leq 1$. Because 
\begin{equation}
    1 \ = \  \operatorname{tr}\left(\rho^{(n)}_{\varrho;\omega}\right) \ = \  \sum_{i=1}^d \lambda_i\left(\rho_{\varrho;\omega}^{(n)}\right)
\end{equation}
holds $\mbQ_{\varrho;\omega}$-almost surely and $M_{\varrho;\omega}^{(2,\infty)}(\bar{a}) = 1$ means that 
\begin{equation}
    \sum_{i=1}^d \lambda_i\left(\rho_{\varrho;\omega}^{(n)}(\bar{a})\right)^2 \ \to \ 1,
\end{equation}
it must be the case that $\lambda_d\bigl(\rho_{\varrho;\omega}^{(n)}(\bar{a})\bigr)\to 1$ and $\lambda_k\bigl(\rho_{\varrho;\omega}^{(n)}(\bar{a})\bigr)\to 0$ for all $k\leq d-1$. That is, $\sigma\bigl(\rho^{(n)}_{\varrho;\omega}(\bar{a})\bigr)$ converges to $\{0, 1\}$ with 1 being a non-degenerate eigenvalue whenever $M_{\varrho;\omega}^{(2)}(\bar{a}) = 1$, as claimed. This, in turn, clearly implies that $M_{\varrho;\omega}^{(m,\infty)}(\bar{a}) = 1$ for all $m\geq 3$. Taking expectation with respect to $\mbQ_{\varrho; \omega}$, we see that 
\begin{equation}
X \ = \  \left\{ 
        \omega\in\Omega\,:\,
        \text{$
                \avg_{\varrho;\omega}\left[
                M_{\varrho;\omega}^{(2,\infty)}
                \right]
                =
                1
                $
            for all $\varrho\in\mbS_d$
            }
    \right\},
\end{equation}
as claimed.
\end{proof}
With the reduction above we see that for any $\varrho\in\mbS_d$, if we let $X_{\varrho} := 
    \left\{ 
       \mbE_{\varrho}\left[M_{\varrho}^{(2,\infty)}\right] = 1
    \right\},$ then the above lemma gives 
\begin{equation}\label{Purification event as an intersection}
    X \ = \  \bigcap_{\varrho\in\mbS_d} X_{\varrho}.
\end{equation}
We claim that $X_\varrho$ is $\mcF$-measurable for all $\varrho\in\mbS_d$. This is straightforward: since $M^{(2, n)}_{\varrho;\omega}\in [0, 1]$, dominated convergence gives 
    \begin{align}
        \avg_{\varrho;\omega}\left[M_{\varrho;\omega}^{(2,\infty)}\right]
        \ = \ 
        \lim_{n\to\infty} \avg_{\varrho;\omega}\left[M_{\varrho;\omega}^{(2, n)}\right].
    \end{align}
But $\omega \mapsto \avg_{\varrho;\omega}\left[M_{\varrho;\omega}^{(2, n)}\right]$ is $\mcF$-measurable since 
\begin{equation}
    \mbE_{\varrho;\omega}\left[M_{\varrho;\omega}^{(2, n)}\right]
        \ = \  
        \sum_{\bar{a}\in\mcA^n}
        \operatorname{tr}\left(V^{(n)}_{\bar{a};\omega} \varrho V^{(n)\dagger}_{\bar{a};\omega} \right)
            M_{\varrho;\omega}^{(2, n)}(\bar{a})
\end{equation}
  Thus $\omega\mapsto \avg_{\varrho;\omega}\left[M_{\varrho;\omega}^{(2,\infty)}\right]$ is also $\mcF$-measurable, being the limit of a sequence of $\mcF$-measurable functions. We record the above discussion in the following lemma.
\begin{lemma}\label{Purification event for specific initial state is measurable}
For any $\varrho\in\mbS_d$, $X_\varrho\in\mcF$. 
\end{lemma}
To finally conclude the measurability of $X$, we prove that the intersection on the right hand side of \eqref{Purification event as an intersection} may be replaced by a countable intersection. To do this, we show that, for fixed $\omega\in\Omega$, the map $\varrho\mapsto \mbE_{\varrho; \omega}\left[M_{\varrho; \omega}^{(2, \infty)}\right]$ is continuous. Towards this end, let $\left\{\Phi^{(n)}_\omega\right\}_{n\in\mbN}$ be the sequence of maps given by 
\begin{align}\label{Uppercase phi}
\begin{split}
    \Phi^{(n)}_\omega :\mbS_d &\to [0,1]\\
    \varrho&\ \mapsto \ \mbE_{\varrho;\omega}\left[M_{\varrho;\omega}^{(2, n)}(\bar{a})\right].
\end{split}
\end{align}
The following proposition, whose proof is given in Appendix \ref{App:Equicontinuity}, establishes the uniform equicontinuity of this sequence of maps. 
\begin{restatable}[]{prop}{Equicontinuity}\label{Equicontinuity}
        For fixed $\omega\in\Omega$, the family of maps $\left\{\Phi_\omega^{(n)}\right\}_{n\in\mbN}$ is uniformly equicontinuous.
\end{restatable}
\begin{proof}
    See Appendix \ref{App:Equicontinuity}.
\end{proof}
A consequence of the above is the following key result. 
\begin{cor}\label{Expectations of second moments are continuous}
For all $\omega\in\Omega$, the map $\mbS_d\to[0, 1]$ given by $\varrho\mapsto \mbE_{{\varrho;\omega}}\left[M^{(2,\infty)}_{\varrho;\omega}\right]$ is continuous.
\end{cor}
\begin{proof}
    Since $\Phi_\omega^{(n)}\in [0, 1]$ for all $n$, the above proposition together with the Arzel\`a-Ascoli theorem give a subsequence $\Phi_\omega^{(n_k)}$ that converges uniformly to some limit, say $\Phi_\omega$. By the previous proposition, each $\Phi_\omega^{(n)}(\cdot)$ is a continuous function, so, being the uniform limit of a sequence of continuous functions on the compact set $\mbS_d$, $\Phi_\omega(\cdot)$ is a continuous function. Moreover, dominated convergence implies that
            \begin{equation}
            \Phi_\omega(\varrho)
            \ = \ 
                \lim_{k\to\infty}\Phi_\omega^{(n_k)}(\varrho)
                \ = \ 
                \lim_{k\to\infty}
               \mbE_{{\varrho;\omega}}\left[M^{(2, n_k)}_{\varrho;\omega}\right]
                \ = \ 
                 \mbE_{{\varrho;\omega}}\left[M^{(2,\infty)}_{\varrho;\omega}\right]
            \end{equation}
            for all $\varrho\in\mbS_d$, i.e., $\varrho\mapsto \mbE_{{\varrho;\omega}}\left[M^{(2,\infty)}_{\varrho;\omega}\right]$ is continuous. 
\end{proof}
Thus, $\omega\in X_\varrho$ for all $\varrho\in\mbS_d$ if and only if $\omega\in X_{\varrho_n}$ for all $n\in\mbN$ where $(\varrho_n)_{n\in\mbN}$ is any countable dense sequence of states in $\mbS_d$. Therefore, we have that 
\begin{equation}
    X \ = \  \bigcap_{n=1}^\infty X_{\varrho_n}
\end{equation}
for any countable dense set $\{\varrho_n:n\in\mbN\}$ of states. From Lemma \ref{Alternative definitions of purification} and Lemma \ref{Purification event for specific initial state is measurable}, we may conclude the following. 
\begin{cor}\label{Purification is an event}
    The purification set $X$ is $\mcF$-measurable. 
\end{cor}
\subsection{Proof of Theorem \ref{0-1 law for purification of disordered quantum trajectories}}\label{Sec3:Proof of 0-1 property}
We are now ready to provide proof of Theorem \ref{0-1 law for purification of disordered quantum trajectories}, which we restate below for convenience.
\firstresult*
\begin{proof}
Corollary \ref{Purification is an event} establishes that $X_\mcV\in \mcF$, so it remains only to prove that $\Pr\left(X_\mcV\right) \in \{0,1\}$. To show this, we prove that $X_\mcV$ is essentially $\theta$-invariant, so that by the ergodicity of $\theta$ we may conclude that $\Pr\left(X_\mcV\right) \in \{0,1\}$.
Let $\omega\in \theta^{-1}(X)$. Then, by the definition of $X$ given in \eqref{Purification event, basic version, for proof of Theorem 1}, for any $\varrho'\in\mbS_d$, 
\begin{equation}\label{Purification event is 0-1, Eqn 1}
\mbQ_{\varrho'; \theta(\omega)}\left(M_{\varrho'; \theta(\omega)}^{(m)} = 1\right) 
     \ = \ 
1.
\end{equation}
So, for arbitrary $\varrho\in\mbS_d$  and all $a\in\mcA$ with $\mbQ_{\varrho; \omega}(\pi_1^{-1}\{a\})>0$, if we let $\varrho(a; \omega)$ denote $V_{{a};{\omega}}\cdot\varrho$, then \eqref{Purification event is 0-1, Eqn 1} with $\varrho' = \varrho(a; \omega)$ gives
\begin{equation}\label{Purification event is 0-1, Eqn 2}
\mbQ_{\varrho(a; \theta(\omega)); \theta(\omega)}\left(M_{\varrho(a; \theta(\omega)); \theta(\omega)}^{(m)} = 1\right) 
     \ = \ 
1.
\end{equation}
Let now $S:\mcA^\mbN\to\mcA^\mbN$ denote the shift $S(a_1, a_2, \dots) = (a_2, a_3, \dots)$. Let $\mbQ_{\varrho;\omega}\left[\ \cdot\ \ \vert \  \pi_1^{-1}\{a\}\right]$ denote the conditional probability (which itself is a probability measure) and $S_*\mbQ_{\varrho;\omega}\left[\ \cdot\ \vert \ \pi_1^{-1}\{a\}\right]$ denote the pushforward of the conditional probability $\mbQ_{\varrho;\omega}\left[\ \cdot\ \ \vert \  \pi_1^{-1}\{a\}\right]$ under $S$. Then
\begin{align}
S_*\mbQ_{\varrho;\omega}\left[\pi_n^{-1}\{(a_1, \dots, a_n)\}\vert\pi_{1}^{-1}\{a\}\right]
    &\ = \  
\frac{\mbQ_{\varrho;\omega}\left(S^{-1}\left(\pi_{n}^{-1}\{(a_1, \dots, a_n)\}\right)\cap \pi_1^{-1}\{a\}\right)}{\mbQ_{\varrho;\omega}\left(\pi_1^{-1}\{a\}\right)} \\
    &\ = \  
\frac{\mbQ_{\varrho;\omega}\left(\pi_{n+1}^{-1}\{(a, a_1, \dots, a_n)\}\right)}{\mbQ_{\varrho;\omega}\left(\pi_1^{-1}\{a\}\right)} \\
    &\ = \ 
\frac{\operatorname{tr}\left(
    V_{(a, a_1, \dots, a_n); \omega}^{(n)}\varrho V_{(a, a_1, \dots, a_n); \omega}^{(n)\dagger}
    \right)}{\operatorname{tr}\left(V_{a;\omega}\varrho V_{a;\omega}^\dagger\right)}
                           \\
    &\ = \  
\tr{
V_{(a_1, \dots, a_n); \theta(\omega)}^{(n)}\varrho(a, \theta(\omega))V_{(a_1, \dots, a_n); \theta(\omega)}^{(n)\dagger}
} \\
    &\ = \ 
\mbQ_{\varrho(a, \theta(\omega));\theta(\omega)}\left(\pi_n^{-1}\{(a_1, \dots, a_n)\}\right)
\end{align}
for any $a_1, \dots, a_n\in\mcA$ and $a\in\mcA$ with $\mbQ_{\varrho;\omega}\left(\pi_1^{-1}\{a\}\right) > 0$. In particular, by a monotone class argument, we have that
\begin{equation}\label{Purification event is 0-1, Eqn 3}
    S_*\mbQ_{\varrho;\omega}\left[ \ \cdot \ \vert \ \pi_{1}^{-1}\{a\}\right]
    \ = \ 
    \mbQ_{\varrho(a; \theta(\omega)); \theta(\omega)}.
\end{equation}
Because
\begin{equation}
    \mbQ_{\varrho; \omega}
    \left(  
        M_{\varrho; \omega}^{(m)}  =   1
    \right)
    \ = \ 
    \sideset{}{'}\sum_{a\in \mcA}
    \mbQ_{\varrho;\omega}\left(\pi_1^{-1}\{a\}\right)
     S_*\mbQ_{\varrho;\omega}\left[M_{\varrho(a; \theta(\omega)); \theta(\omega)}^{(m)} = 1\,\big\vert\,\pi_{1}^{-1}\{a\}\right] \ ,
\end{equation}
where $\sideset{}{'}\sum$ denotes the sum over $a\in \mcA$ such that $\mbQ_{\varrho;\omega}\bigl(\pi_1^{-1}\{a\}\bigr) > 0$,
from \eqref{Purification event is 0-1, Eqn 1} and \eqref{Purification event is 0-1, Eqn 3} we have 
\begin{align}
     \mbQ_{\varrho; \omega}
    \left(  
        M_{\varrho; \omega}^{(m)} = 1
    \right)
    &\ = \ 
    \sideset{}{'}\sum_{a\in \mcA}
    \mbQ_{\varrho;\omega}\left(\pi_1^{-1}\{a\}\right)
     \mbQ_{\varrho(a; \theta(\omega)); \theta(\omega)}\Big(M_{\varrho(a; \theta(\omega)); \theta(\omega)}^{(m)} = 1\Big) \\
     &\ = \  
     \sideset{}{'}\sum_{a\in \mcA}
    \mbQ_{\varrho;\omega}\left(\pi_1^{-1}\{a\}\right) \ = \  1
\end{align}
whereby $\omega\in X$. This implies that $X$ is essentially $\theta$-invariant. As mentioned above, the ergodicity of $\theta$ allows us to conclude $\Pr(X)\in\{0, 1\}$. 
\end{proof}

\section{Dark subspaces}\label{Sec:Dark subspaces}
The goal of this section is to develop the theory of disordered dark subspaces and to prove Theorem \ref{Main theorem'}. As a corollary of Theorem \ref{Main theorem'}, Theorem \ref{Main theorem} is also proved. 
\subsection{Preliminaries and notation}
We generalize the notion of dark subspaces to the disordered setting. Along the way, we introduce the notion of \emph{gray subspaces} and recall the definition of random compact sets. 
Recall that $\scrP$ denotes the set of projections in $\matrices$ and that $\scrP_r$ denotes the rank $r$ projections. For a subset $S\subset\matrices$, we let $\scrP(S)$ denote the set of projections in $S$, and, for $r\in\mbN$, we let $\scrP_r(S)$ denote the subset of $\scrP(S)$ consisting of rank $r$ projections. Let $\mcK(\matrices)$ denote the collection of compact subsets of $\matrices$. Recall that a map $\Omega\to\mcK(\matrices)$ is said to be \emph{a random compact set} if it is measurable with respect to the Borel $\sigma$-algebra generated by the Hausdorff metric on $\mcK(\matrices)$. 
Given a disordered perfect Kraus measurement ensemble $\mcV$ on $\mcA$, the set $\scrD_\omega$ was defined to be the set of those $p\in\scrP$ such that 
\begin{equation}
    pV_{\bar{a}; \omega}^{(n) \dagger}V_{\bar{a}; \omega}^{(n)} p \ \propto \ p
\end{equation}
for all $n\in\mbN$ and $\bar{a}\in\mcA^n$, and $\scrD_\omega^{(r)}$ was defined as $\scrD_\omega\cap\scrP_r$. In this section, we establish some preliminary facts about the association $\omega\mapsto\scrD^{(r)}_\omega$. Namely, that this association gives a random compact set. From \cite[Proposition 1.1.2]{molchanov2017theory} and \cite[Theorem 1.3.14 (iii)]{molchanov2017theory}, a compact set-valued map $K: \Omega\to\mcK(\matrices)$ is measurable with respect to the Hausdorff metric if and only if the following, easier-to-work with definition for measurability is satisfied. We further refer to the reader to  \cite[\S 18]{guide2006infinite} and \cite{kuratowski1965general,
 rockafellar1969measurable} for results on measurable set-valued functions, also known as measurable correspondences. 
\begin{definition}\label{Dark subspaces: Definition of random set}
    A map $\Omega\to\mcK(\matrices)$ given by $\omega\mapsto \scrX_\omega$ is called \emph{measurable} if for every compact set $K\in\mcK(\matrices)$, it holds that 
    \begin{equation}
        \{\omega\in\Omega\,:\,\scrX_\omega\cap K\neq\emptyset\}
    \end{equation}
    is a measurable set. A measurable map $\omega\mapsto \scrX_\omega$ is referred to as a \emph{random compact set}, or simply a random set when compactness is clear. 

\end{definition}  
As a first step towards showing $\omega\mapsto\scrD_\omega^{(r)}$ is measurable, we make the following definition.
\begin{definition}[Gray projections]\label{Dark subspaces: n-dark projections of rank r definition}
    Let $\mcV$ be a disordered perfect Kraus measurement on $\mcA$ and fix $\omega\in\Omega$. For $N,r\in\mbN$, let 
    \begin{equation}\label{Dark subspaces: n-dark projections of rank r definition, equation}
        \scrG_\omega^{(N, r)}
        \ := \  \bigcap_{\bar{a}\in\mcA^k}\left\{
            p\in\scrP_r\,:\, \text{there is $\lambda\in[0, 1]$ such that }pV_{\bar{a};\omega}^{(N)\dagger}V_{\bar{a};\omega}^{(N)}p= \lambda p
        \right\}.
    \end{equation}
    We refer to $\scrG^{(N, r)}_\omega$ as the set of \emph{$N$-gray projections of rank $r$} for $\{V_{a;\omega}\}_{a\in\mcA}$. The union $\bigcup_{r=2}^d\scrG_\omega^{(N, r)}$ is denoted $\scrG_\omega^{(N)}$ and called the set of \emph{$N$-gray projections}. The range of a gray projection is called a \emph{gray subspace}. 
\end{definition}
Thus, the set of dark projections of rank $r$ may be expressed as the intersection 
\begin{equation}\label{Dark subspaces of rank r as intersection of gray subspaces of rank r}
     \scrD_\omega^{(r)} \ = \ \bigcap_{N\in\mbN} \scrG_\omega^{(N, r)}
\end{equation}
and the set $\scrD_\omega$ of all dark projections of rank $r\geq 2$ may be expressed as 
\begin{equation}\label{Dark subspaces as intersection of gray subspaces}
    \scrD_\omega \ = \ \bigcap_{N\in\mbN}\scrG_\omega^{(N)} \ .
\end{equation}
We re-state this as a formal definition of the set of dark projections:
\begin{definition}[Dark projections]\label{Dark subspaces: dark projections}
    Let $\mcV$ be a disordered perfect Kraus measurement on $\mcA$ and fix $\omega\in\Omega$. For $N,r\in\mbN$, given a disorder realization $\omega\in\Omega$ we denote by $\scrD_\omega$ the associated set of \emph{dark projections}, which is defined as
        \begin{equation}
            \scrD_\omega \ = \ \bigcap_{N\in\mbN}\scrG_\omega^{(N)} \ ,
        \end{equation}
    where $\scrG_\omega^{(N)}$ are the $N$-gray projections as defined in Definition \ref{Dark subspaces: n-dark projections of rank r definition}. 
\end{definition}
Now, we turn to establishing basic properties of gray subspaces. 
\begin{lemma}\label{Dark subspaces: The set of N-gray subspaces forms a nested sequence}
    For all $N\in\mbN$ and almost every $\omega\in\Omega$, $\scrG_\omega^{(N+1)}\subset\scrG_\omega^{(N)}$.
\end{lemma}
\begin{proof}
    Let $p\in\scrG_\omega^{(N+1)}$ and fix $\bar{a}\in\mcA^{N}$. Then for each $a\in \mcA$ there is $\lambda_{(\bar{a}, a)}\in[0, 1]$ such that
    \begin{equation}
        pV_{\bar{a}; \omega}^{(N)\dagger}V_{a; \theta^{N+1}(\omega)}^\dagger V_{a; \theta^{N+1}(\omega)}
        V_{\bar{a}; \omega}^{(N)}p
        \ = \
        \lambda_{(\bar{a}, a)} p \ .
    \end{equation}
   Since $\mcV_{\theta^{N+1}(\omega)}\in\scrV_\mcA$, summing over $a\in\mcA$ yields 
    \begin{equation}\label{Dark subspaces: The set of N-gray subspaces forms a nested sequence, Eqn 1}
         pV_{\bar{a}; \omega}^{(N)\dagger}
        V_{\bar{a}; \omega}^{(N)}p
        \  = \ 
        \left(\sum_{a\in\mcA}\lambda_{(\bar{a}, a)}\right)p \ .
    \end{equation}
    Setting $\mu_{\bar{a}} = \sum_{a\in\mcA}\lambda_{(\bar{a}, a)}$, it is clear that $\mu_{\bar{a}}\geq 0$, and $\mu_{\bar{a}}\leq 1$ since $\mcV\in\scrV_\mcA$ implies 
    \begin{equation}
        1 \  = \  \sum_{a\in\mcA}\sum_{\bar{a}\in\mcA^N}\lambda_{(\bar{a}, a)} \  = \  \sum_{\bar{a}}\mu_{\bar{a}} \ .
    \end{equation}
    Thus $p\in\scrG_\omega^N$. 
\end{proof}
\begin{lemma}\label{Dark subspaces: The set of N-gray subspaces of rank r is compact}
    For all $\omega\in\Omega$, the set $\scrG_\omega^{(N, r)}$ of $N$-gray projections of rank $r$ is a compact subset of $\matrices$. 
\end{lemma}
\begin{proof}
    Because orthogonal projections have operator norm 1, it suffices to show that $\scrG_\omega^{(N, k)}$ is closed. Moreover, because of the expression of $\scrG_\omega^{(N, k)}$ as an intersection of sets, it suffices to note that  
    \begin{equation}
        G_{\bar{a}} \ := \ \left\{
            p\in\scrP_r\,:\, \text{there is $\lambda\in[0, 1]$ such that }p V_{\bar{a};\omega}^{(N)\dagger}V_{\bar{a};\omega}^{(N)} p = \lambda p
        \right\}
    \end{equation}
    is closed for each $\bar{a}\in \mcA^N$. To see this, suppose that $(p_n)_{n\in\mbN}$ is convergent sequence in $G_{\bar{a}}$.  Since $\scrP_r$ is closed, it is immediate that $p=\lim_n p_r\in \scrP_r$.  Furthermore, for each $n\in\mbN$, there is $\lambda_n\in [0, 1]$ with $p_n V_{\bar{a};\omega}^{(N)\dagger} V_{\bar{a};\omega}^{(N)} p_n = \lambda_n p_n$. By passing to a sub-sequence, $(\lambda_n)_{n\in \mbN}$ is convergent.  It follows that $p V_{\bar{a};\omega}^{(N)\dagger} V_{\bar{a};\omega}^{(N)} p = \lim_n\lambda_n p$, and thus that $p\in G_{\bar{a}}$ as desired.
\end{proof}
As a corollary to the previous two lemmas and the expression \eqref{Dark subspaces as intersection of gray subspaces}, we get the following technical result that will prove useful below. 
\begin{cor}\label{Dark subspaces: Sufficient condition for dark subspaces nonempty}
    A necessary and sufficient condition for $\scrD_\omega$ to be nonempty is that $\scrG_\omega^{(N)}$ is nonempty for all $N\in\mbN$. 
\end{cor}
\begin{remark}\label{rem:technicalconstructive}
    Theorem \ref{Main theorem'} and the contrapositive of this result imply the following:
    \begin{quote}\emph{For asymptotic purification to hold it is necessary and sufficient that  $\Pr[\scrG_\omega^{(N)}=\emptyset]>0$ for some $N>0$.}  
    \end{quote}
    This is a more precise statement of the constructive criterion for asymptotic purification mentioned in Remark \ref{rem:constructive}. Note that the proof of asymptotic purification in Example \ref{Example:Harder to prove purifying example} amounted to showing $\scrG_\omega^{(2)}=\emptyset$ with probability one. 
\end{remark}
We now address the issue of measurability. 
\begin{lemma}\label{Dark subspaces: N-gray projections of rank r is a random set}
    For all $k\in\mbN$ and $\bar{a}\in \mcA^k$, the map
    \begin{equation}\label{Dark subspaces: N-gray projections of rank r is a random set, Eqn 1}
        \omega\ \mapsto\  \{p\in\scrP_r\,:\, \text{there is $\lambda\in[0, 1]$ with $p V_{\bar{a};\omega}^{(k)\dagger} V_{\bar{a};\omega}^{(k)} p = \lambda p$}\}
    \end{equation}
    is measurable. 
\end{lemma}
\begin{proof}
    As above, denote the set on the r.h.s.\ of \eqref{Dark subspaces: N-gray projections of rank r is a random set, Eqn 1} by $G_{\bar{a}}$. The proof of the previous lemma established the compactness of $G_{\bar{a}}$. So, by Definition \ref{Dark subspaces: Definition of random set}, it suffices to show the set
    \begin{equation}
        \left\{
            \omega\in\Omega\,:\, 
                G_{\bar{a}}\cap K\neq\emptyset 
        \right\}
        \  = \  
        \left\{
            \omega\in\Omega\,:\, 
                \text{$\exists$ $p\in \scrP_r(K)$ and $\lambda\in[0, 1]$ such that $p V_{\bar{a};\omega}^{(k)\dagger}V_{\bar{a};\omega}^{(k)} p = \lambda p$}
        \right\}
    \end{equation}
    is measurable for any $K\in\mcK(\matrices)$. This is equivalent to asking that the function 
    \begin{equation}
        \omega\mapsto 
        1_{
        \left\{
            V\in \matrices\,:\, \text{$\exists$ $p\in \scrP_r(K)$ and $\lambda\in[0, 1]$ such that $p V^\dagger V p = \lambda p$}
        \right\}
        }
        \left(
        V_{\bar{a};\omega}^{(k)}
        \right)
    \end{equation}
    be measurable. Because $\omega\mapsto V_{\bar{a};\omega}^{(k)}$ is measurable, we just need to show that 
    \begin{equation}
        \mcW
    \ := \ 
    \left\{
            V\in \matrices\,:\, \text{$\exists$ $p\in \scrP_r(K)$ and $\lambda\in[0, 1]$ such that $p V^\dagger V p  = \lambda p$}
        \right\}
    \end{equation}
    is a Borel set. In fact, $\mcW$ is closed, as can be verified by a convergence argument similar to that in the proof of Lemma \ref{Dark subspaces: The set of N-gray subspaces of rank r is compact}.   
\end{proof}
It follows from Definition \ref{Dark subspaces: Definition of random set} that countable intersections and finite unions of random compact sets are themselves random compact sets, so that by the definitions of $\scrG^{(N, r)}$ and $\scrD^{(r)}$, we have the following corollary. 
\begin{cor}\label{Dark subspaces: Gray and dark subspaces are random sets}
    The maps $\omega\mapsto\scrG_\omega^{(N, r)}$, $\omega\mapsto\scrG_\omega^{(N)}$, $\omega\mapsto\scrD_\omega^{(r)}$, and $\omega\mapsto \scrD_\omega$ are $\mcF$-measurable. 
\end{cor}
In light of previous corollary, it follows from the so-called Kuratowski–Ryll-Nardzewski measurable selection theorem \cite[Proposition 1.6.3]{arnold1998random}, \cite{kuratowski1965general,aumann1965integrals}, that measurable sections of $\scrD_\omega$ exists.
\begin{cor}\label{Dark subspaces: There exist measurable fibers}
    Whenever $\scrD_\omega$ is almost surely nonempty, there exists a measurable function $\varphi:\Omega\to\scrP$ such that $\varphi(\omega)\in\scrD_\omega$ almost surely. 
\end{cor}
In the following, we demonstrate that $\scrD_\omega$ is almost surely nonempty in the situation that $\mbP(X) = 0$. Thus, the above corollary implies that there is a measurable manner in which we may select dark subspaces given that $\mcV$ does not asymptotically purify. 
\subsection{Existence of dark subspaces}
Our present goal is to prove the following proposition.
\begin{prop}\label{Main theorem, technical part}
    Let $X$ denote the purification event. For every $N\in\mbN$ and almost every $\omega\in \Omega \setminus X$, the set $\scrG_\omega^{(N)}$ is nonempty. 
\end{prop}
The proof of this proposition proceeds roughly as \cite[proof of Lemma 3]{maassen2006purification}. However we note that due to the presence of the ``shifts" $\theta^{n}$ in various expressions, the precise techniques of the proof in \cite{maassen2006purification} do not work in the disordered case. This can be seen by contrasting Lemma \ref{Proof of main theorem: Technical lemma 2} and Proposition \ref{Proof of main theorem: Technical lemma 3} with to \cite[Lemma 3]{maassen2006purification}. 
\subsubsection{Submartingale increments}\label{Subsec:Submartingale incremenets}
For $\omega\in\Omega$ we introduce a non-negative function on the set of states $\mbS_D$ as follows. Let $m\in\mbN$ and $\omega\in\Omega$, and define 
\begin{align}
    \begin{split}
        \delta^{(m)}_{\omega}: \mbS_d&\to \mbR_{\geq 0}\\
        \varrho&\ \mapsto \ \sum_{a\in\mcA}
                                        \operatorname{tr}\big(V_{a;\omega}\varrho V_{a;\omega}^{\dagger}\big)
                                        \left(
                                                \operatorname{tr}\left(\left(V_{a;\omega}\cdot\varrho\right)^m\right)
                                                -
                                                \operatorname{tr}\left(\varrho^m\right)
                                        \right)^2.
    \end{split}
    \label{delta_m definition}
\end{align}
Note that  $(\varrho, \omega)\mapsto \delta^{(m)}_{\omega}(\varrho)$ is $(\mcB(\mbS_d)\otimes\mcF)$-measurable, and that $\delta^{(m)}$ is continuous in $\varrho$. Moreover, for any $n\in\mbN$, $\varrho\in\mbS_d$, and $\omega\in\Omega$, 
\begin{align}
 \delta^{(m)}_{\theta^{n+1}(\omega)}\left(\rho_{\varrho;\omega}^{(n)}\right)
                                    &\  = \ \sum_{a\in\mcA}
                                        \operatorname{tr}\left(V_{a;\theta^{n+1}(\omega)}\rho^{(n)}_{\varrho;\omega} V_{a;\theta^{n+1}(\omega)}^\dagger\right)
                                        \left(
                                        \tr{
                                        \left(
                                                    V_{a;\theta^{n+1}(\omega)}\cdot\rho^{(n)}_{\varrho;\omega}
                                                \right)^m
                                        }
                                                -
                                                \operatorname{tr}\left(\rho^{(n)\,\,m}_{\varrho;\omega}\right)
                                        \right)^2\notag\\
                                    &\  = \  \mbE_{{\varrho;\omega}}
                                        \left[
                                            \left(M_{\varrho;\omega}^{(m, n+1)} - M_{\varrho;\omega}^{(m, n)}
                                            \right)^2
                                            \,\Big\vert\,
                                            \Sigma_n
                                        \right]
        \label{delta_m as martingale increments}
\end{align}
holds $\mbQ_{\varrho; \omega}$-almost surely. Now, since $\left(M_{\varrho; \omega}^{(m, n)}\right)_{n\in\mbN}$ is a submartingale taking values in $[0, 1]$, its increments are square summable. So, \eqref{delta_m as martingale increments} gives
\begin{equation}\label{summability}
     \sum_{n=0}^\infty 
     \mbE_{{\varrho;\omega}}
     \left[\delta^{(m)}_{\theta^{n+1}(\omega)}\left(\rho_{\varrho;\omega}^{(n)}\right)\right] \ \leq \ 1 \ .
\end{equation}
In particular, 
\begin{equation}\label{limitzero}
        \lim_{n\to\infty}
        \mbE_{{\varrho;\omega}}\left[\delta^{(m)}_{\theta^{n + 1}(\omega)}\left(\rho_{\varrho;\omega}^{(n)}\right)\right]
        \  = \  0 \ .
\end{equation}

Note that the above argument worked pointwise a.s. in $\omega$ and is easily adapted to give \eqref{summability} \eqref{limitzero} with $\varrho$ a random initial state, depending on $\omega$. We record this as a lemma. 
\begin{lemma}\label{Proof of main theorem: Technical lemma 1}Let $\varrho:\Omega\to \mbS_d$ be a random state.  Then 
\begin{equation} \sum_{n=0}^\infty 
     \mbE_{{\varrho;\omega}}
     \left[\delta^{(m)}_{\theta^{n+1}(\omega)}\left(\rho_{\varrho_\omega;\omega}^{(n)}\right)\right] \ \leq \ 1 \quad \text{and} \quad  \lim_{n\to\infty}
        \mbE_{{\varrho;\omega}}\left[\delta^{(m)}_{\theta^{n + 1}(\omega)}\left(\rho_{\varrho_\omega;\omega}^{(n)}\right)\right]
        = 0 \quad \text{a.s.} \ .
        \end{equation}
\end{lemma}

\subsubsection{Measurability considerations}\label{Subsec:Measurability considerations}
We now state and prove a non-deterministic version of \cite[Lemma 3]{maassen2006purification}. We note that several technical issues arise in the present random setting, for example the question of measurability for the $\mbS_d$-valued sequence of random variables $(\bar{\varrho}_n)_{n\in\mbN}$ in the following Lemma.
\begin{lemma}\label{Proof of main theorem: Technical lemma 2}
Let $X$ denote the purification event. Then there is a random state $\phi:\Omega \to \mbS_d$ such that for   
        \begin{equation}\label{eq:mu}
            \mu(\omega) \ := \  \avg_{\phi(\omega);\omega}\left[M^{(2,\infty)}_{\phi(\omega);\omega}\right] \ ,
        \end{equation}
         we have $\mu<1$ almost surely on $\Omega\setminus X$. Moreover, for all $n$ there is a random state $\bar\varrho_n:\Omega\to\mbS_d$ satisfying the following conditions. 
            \begin{enumerate}
                \item[(i)] For almost every $\omega \in \Omega$, 
                \begin{equation}\label{eq:technical lemma 2 i}
                    \tr{\bar\varrho_n(\omega)^2}\leq \dfrac{\mu(\omega)+1}{2} \ .
                \end{equation}

                \item[(ii)] For almost every $\omega \in \Omega\setminus X$,  
                \begin{equation}\label{eq:technical lemma 2 ii}
                    \sum_{m=1}^d\delta_{\theta^{n+1}(\omega)}^{(m)} (\bar\varrho_n(\omega)) \leq \sum_{m=1}^d\frac{1+\mu(\omega)}{1-\mu(\omega)} \mbE_{{\phi(\omega)};\omega}\left[\delta^{(m)}_{\theta^{n+1}(\omega)}\left(\rho^{(n)}_{\phi(\omega);\omega}\right)\right] \ .
                \end{equation}
            \end{enumerate}
\end{lemma}
\begin{proof}
      Let $(\phi_n)_{n\in\mbN}$ be any countable dense set of states on $\mbS_D$. Then from Proposition \ref{Equicontinuity} and Lemma \ref{Alternative definitions of purification} we have that 
      \begin{equation}
          X \  = \  \bigcap_{n\in\mbN} \left\{\omega: \avg_{\phi_n;\omega}\left[M^{(2,\infty)}_{\phi_n;\omega}\right] = 1\right\}\ .
      \end{equation}
      Now we define a random function $\tau:\Omega \to \mbN\cup\{\infty\}$ such that 
        \begin{equation}
            \tau(\omega) \  = \  \inf\left\{n\in\mbN : \avg_{\phi_n;\omega}\left[M^{(2,\infty)}_{\phi_n;\omega}\right] < 1\right\}
        \end{equation}
        where the infimum of the empty set is defined to be $\infty$. We note that $\tau=\infty$ on the purification event $X$ and that $\tau<\infty$ on $\Omega\setminus X$. Furthermore $\tau$ is an $\mcF$-measurable map, since the map $\omega\mapsto \avg_{\varrho;\omega}\left[M^{2,\infty}_{\varrho;\omega}\right]$ is $\mcF$-measurable for any initial $\varrho$. 
            
        We define the random state $\phi$ as follows. 
            \begin{equation}
            \phi(\omega) \  = \  
                \begin{cases}
                    \phi_n & \text{if } \tau(\omega) = n\\
                    \phi_1 & \text{if } \tau(\omega) = \infty. 
                 \end{cases}
            \end{equation}
        It follows from Lemma \ref{stopped random elememt} below that $\omega \mapsto\phi(\omega)$ is $\mcF$-measurable.  Furthermore, with $\mu$ as in \eqref{eq:mu}, we see that $\mu$ is $\mcF$-measurable (again by Lemma \ref{stopped random elememt}, $\mu=1$ on $X$ and $\mu<1$ on $\Omega\setminus X$. 

        It remains to construct the sequence $(\bar \varrho_n)_{n=1}^\infty$ of random states satisfying conditions (i) and (ii) of the lemma.  We start by observing that for each $n\in\mbN$, the map $(\omega,\bar{a})\mapsto M^{(2,n)}_{\phi(\omega);\omega}(\bar{a})$ is $\mcF\otimes\Sigma_n$-measurable, by another application of Lemma \ref{stopped random elememt}. Therefore, for each $k$ the following set is $\mcF\otimes\Sigma_k$-measurable:
            \begin{equation}
                A^{(k)} \  := \  \left\{(\omega,\bar{a}) : M^{(2,k)}_{\phi(\omega);\omega}(\bar{a}) \leq \dfrac{\mu(\omega)+1}{2}\right\} \ .
            \end{equation}
        By Tonelli's theorem the the $\omega$-section of $A^{(k)}$, 
            \begin{equation}
                A^{(k)}_\omega \  = \  \left\{\bar{a} : (\omega,\bar{a}) \in A^{(k)}\right\} \ ,
            \end{equation}
        is $\Sigma_k$-measurable for almost every $\omega$. We claim further, that for almost every $\omega \in \Omega \setminus X$, the set $A^{(k)}_\omega$ has positive $\mbQ_{\phi(\omega);\omega}$-measure. To see this, note that 
            \begin{equation}
                \mu(\omega) \  = \  \avg_{\phi(\omega);\omega}\left[M^{(2,\infty)}_{\phi(\omega);\omega}\right]\ 
                \geq \ \avg_{\phi(\omega);\omega}\left[M^{(2,k)}_{\phi(\omega);\omega}\right],
            \end{equation}
        where the inequality follows from the sub-martingale property in Proposition \ref{Submartingale convergence}. Therefore, 
            \begin{equation}
                \mu(\omega) \ \geq \ \avg_{\phi(\omega);\omega}\left[M^{(2,k)}_{\phi(\omega);\omega} \cdot 1_{\mcA^\mbN\setminus(A^{(k)}_\omega)}\right] \ \geq \ \left(\dfrac{\mu(\omega) +1}{2}\right) \mbQ_{\phi(\omega) ;\omega}\left(\mcA^\mbN\setminus A^{(k)}_\omega\right) \ ,
            \end{equation}
        which implies that 
            \begin{equation}
                \mbQ_{\phi(\omega);\omega}\left(A^{(k)}_\omega\right) \ \geq \ \dfrac{1-\mu(\omega)}{1+\mu(\omega)} \ . 
            \end{equation}
        In particular we have  $\mbQ_{\phi(\omega);\omega}\left(A^{(k)}_\omega\right)>0$ for almost every $\omega\in \Omega\setminus X$. 

        Since $\rho^{(k)}_{\phi(\omega);\omega}(\bar{a})$ depends only on the first $k$ observed outcomes $a_1,a_2\ldots a_k$, it follows that  $\rho^{(k)}_{\phi(\omega);\omega}(\bar{a})$ is constant on $\pi_k^{-1}(a_1,a_2\ldots a_k)$. We denote this value of $\rho^{(k)}_{\phi(\omega);\omega}(\bar{a})$ by $\rho^{(k)}_{\phi(\omega);\omega;(a_1\ldots a_k)}$ and note that $\omega\mapsto \rho^{(k)}_{\phi(\omega);\omega;(a_1\ldots a_k)}$ is $\mcF$-measurable. 

        Let $k\in\mbN$ and let $\{b_1,b_2\ldots b_{|\mcA^k|}\}$ be an enumeration of the points of $\mcA^k$. Let $a\in\mcA$ and define $\tilde b_i = (b_i,a,a\ldots) \in \mcA^\mbN$.  Then define the random map $\sigma:\Omega \to \mbN\cup\{\infty\}$ by $\sigma = \max (\sigma_1, \sigma_2)$, where
            \begin{equation}
                \sigma_1(\omega) \ := \  \inf\left\{i: \tilde b_i \in A^{(k)}_\omega\right\}
            \end{equation}
            and 
            \begin{equation}
                \sigma_2(\omega) \ := \ 
                    \inf\left\{i: \sum_{m=1}^d \delta^{(m)}_{\theta^{k+1}(\omega)}\left(\rho^{(k)}_{\phi(\omega);\omega}(\tilde b_i)\right)\leq\dfrac{1+\mu(\omega)}{1-\mu(\omega)} \avg_{\phi(\omega);\omega}\left[\sum_{m=1}^d \delta^{(m)}_{\theta^{k+1}(\omega)}\left(\rho^{(k)}_{\phi(\omega);\omega)}\right)\right] \right\} \ ,
            \end{equation} 
        where we take $\inf \emptyset=\infty$ and further $\sigma_2(\omega)=\infty$ if $\mu(\omega)=1$. We claim that $\sigma(\omega)<\infty$ for almost every $\omega\in \Omega\setminus X$. Firts, for $\omega\in \Omega\setminus X$, $\sigma_1(\omega)$ is finite since then $A^{(k)}_\omega$ has positive $\mbQ_{\phi(\omega);\omega}$-measure. But now for $\omega \in \Omega\setminus X$ and $m\in\{1,\ldots d\}$, we also have 
            \begin{equation}
                \avg_{\phi(\omega);\omega}\left[\delta^{(m)}_{\theta^{k+1}(\omega)}\left(\rho^{(k)}_{\phi(\omega);\omega}\right)\right] \ \geq \ \sideset{}{'}\sum_{\bar{a}\in \mcA^k} \delta^{(m)}_{\theta^{k+1}(\omega)}\left(\rho^{(k)}_{\phi(\omega);\omega}(\bar{a})\right)\mbQ_{\phi(\omega);\omega}\left(\{\bar a \} \times \mcA\times \mcA\ldots \right)
            \end{equation}
        where $\sum'$ denotes the sum over the points $\bar{a}\in \mcA^k$ such that $\pi^{-1}_k(\bar{a}) \subset A^{(k)}_\omega$. Therefore, 
            \begin{multline}
                \dfrac{1}{\mbQ_{\phi(\omega);\omega}\left(A^{(k)}_\omega\right)}\sideset{}{'}\sum_{\bar{a}\in \mcA^k}\mbQ_{\phi(\omega);\omega}\left(\{\bar{a}\} \times \mcA\times \mcA\ldots \right)\left(\sum_{m=1}^d \delta^{(m)}_{\theta^{k+1}(\omega)}\left(\rho^{(k)}_{\phi(\omega);\omega}(\bar{a})\right)\right)\\
                \leq \ \dfrac{1+\mu(\omega)}{1-\mu(\omega)}\sum_{m=1}^d \avg_{\phi(\omega);\omega}\left[\delta^{(m)}_{\theta^{k+1}(\omega)}\left(\rho^{(k)}_{\phi(\omega);\omega}\right)\right] \ .
            \end{multline}
        However, this can happen only if there is some $b_i$ in $\mcA^k$ such that $\tilde b_i \in A^{(k)}_\omega$ and 
            \begin{equation}
                \sum_{m=1}^d \delta^{(m)}_{\theta^{k+1}(\omega)}\left(\rho^{(k)}_{\phi(\omega);\omega}(\tilde b_i) \right) \ \leq \  \dfrac{1+\mu(\omega)}{1-\mu(\omega)}\sum_{m=1}^d \avg_{\phi(\omega);\omega}\left[\delta^{(m)}_{\theta^{k+1}(\omega)}\left(\rho^{(k)}_{\phi(\omega);\omega}\right)\right] \ .
            \end{equation}
        Thus $\sigma_2(\omega)$ is also finite almost surely on $\Omega\setminus X$. Furthermore, $\sigma$ is measurable, as can be seen similar to the proof of measurability of $\tau$.  Defining, for each  $k\in\mbN$,
            \begin{equation}
                \bar\varrho_k(\omega) \  := \  \rho^{(k)}_{\phi(\omega);\omega}(b_{\sigma(\omega)})
            \end{equation}
        where $b_{\sigma(\omega)}$ is defined to be $b_1$ whenever $\sigma(\omega)=\infty$, we see that $\omega \mapsto \bar\varrho_k(\omega)$ is measurable by another application of Lemma \ref{stopped random elememt} and is thus a sequence of random states with the desired properties. 
\end{proof}

\begin{prop}\label{Proof of main theorem: Technical lemma 3}
   For almost every $\omega\in \Omega\setminus X$ there is a mixed state $\psi$ satisfying $\delta^{(m)}_\omega(\psi) = 0$ for all $m\in\{1,2\ldots d\}.$ 
\end{prop}
\begin{remark} The mixed state $\psi$, of course, depends on $\omega$. However, we are not asserting the measurability of the mapping $\omega \mapsto \psi$, since the proof given below does not show this. With some additional effort, one could show that $\mcN_\omega=\{\psi \ : \delta_\omega^{(m)}(\psi)=0, \ m=1,\ldots,d\}$ is a measurable set that is non-empty almost surely on and obtain the existence of a measurable section $\omega \mapsto \psi_\omega$ from the Kuratowski–Ryll-Nardzewski measurable selection theorem \cite[Proposition 1.6.3]{arnold1998random}.
\end{remark}
\begin{proof}
    If $\Pr(X)=1$, there is nothing to prove.  Thus suppose $\Pr(\Omega\setminus X)=1$.
    Let $\phi$, $\mu$ and $(\bar\varrho_n)_{n\in\mbN}$ be as in the previous lemma. Applying powers of the m.p.t.\ $\theta$ to \eqref{eq:technical lemma 2 i} and \eqref{eq:technical lemma 2 ii}, we see that, for all $n\in\mbN$,
    \begin{equation}\label{subset of mixed states}
        \tr{\bar\varrho_n(\theta^{-n-1}(\omega)}^2 \ \leq \ \dfrac{1+\mu(\theta^{-n-1}(\omega))}{2} \quad \text{a.s.}
    \end{equation}
    and
    \begin{multline}\label{inverse shifted delta}
        \displaystyle\sum_{m=1}^d\delta_{\omega}^{(m)} (\bar\varrho_n(\theta^{-n-1}(\omega))) \\ \leq  \ \frac{1+\mu(\theta^{-n-1}(\omega))}{1-\mu(\theta^{-n-1}(\omega))} \sum_{m=1}^d\mbE_{{\phi(\theta^{-n-1}(\omega))};\theta^{-n-1}(\omega)}\left[\delta^{(m)}_{\omega}\left(\rho^{(n)}_{\phi(\theta^{-n-1}(\omega));\theta^{-n-1}(\omega)}\right)\right]\quad \text{a.s.} \ .
    \end{multline}
    Similarly, from Lemma \ref{Proof of main theorem: Technical lemma 1}, we obtain
    \begin{align}\label{integration w.r.t disorder is finite}
                1 \ &\ge \
                \int_{\Omega} \ d\Pr(\omega) \ \sum_{n=0}^\infty \avg_{\phi(\omega);\omega}\left[\delta^{(m)}_{\theta^{n+1}(\omega)}\left(\rho^{(n)}_{\phi(\omega);\omega} \right)\right] \nonumber
                \\
                &\  = \  \int_{\Omega} \ d\Pr(\omega)  \
                \sum_{n=0}^\infty \avg_{\phi(\theta^{-n-1}(\omega));\theta^{-n-1}(\omega)}\left[\delta^{(m)}_{\omega}\left(\rho^{(n)}_{\phi(\theta^{-n-1}(\omega));\theta^{-n-1}(\omega)} \right)\right] \ .
    \end{align}
    Thus 
    \begin{equation}\label{sumbale imlies 0 limit}
                \lim_{n\to\infty} \avg_{\phi(\theta^{-n-1}(\omega));\theta^{-n-1}(\omega)}\left[\delta^{(m)}_{\omega}\left(\rho^{(n)}_{\phi(\theta^{-n-1}(\omega));\theta^{-n-1}(\omega)} \right)\right] \to 0 \quad \text{a.s.} \ .
    \end{equation}

    Since $\Pr(X)=0$, we have $\mu <1$ almost surely and $\lim_n \Pr(E_n)=1$ with $E_n =\{\omega: \mu(\omega) < 1-1/n\}$. In particular there is $n_0$ such that $\Pr(E_{n_0})>0$. By Birkhoff's ergodic theorem, $\Pr(E_{n_0})= \lim_N \sum_{n=1}^N 1_{E_{n_0}}[\theta^{-n}(\omega)]$ for almost every $\omega$. Thus, the sequence $\theta^{-k}(\omega)$ visits $E_{n_0}$ infinitely often. Henceforth we fix a full measure event $\Xi$ on which 1) $(\theta^{-n}\omega)_{n=1}^\infty$ visits $E_{n_0}$ infinitely often and  2) \eqref{subset of mixed states}, \eqref{inverse shifted delta}, and \eqref{sumbale imlies 0 limit} hold. 

    Fix $\omega\in \Xi$.  The proof will be completed by finding a $\psi$ satisfying $\delta^{(m)}_\omega(\psi)$ for $m=1,\ldots,d$. Let $(n_k)_{k=1}^\infty$ be a subsequence such that $\theta^{-n_k-1}(\omega)\in E_{n_0}$ and let 
        \begin{equation}
            \varphi_k \ := \  \bar\varrho_{n_k}(\theta^{-n_k-1}(\omega)) \ .
        \end{equation}
    From \eqref{subset of mixed states} and \eqref{inverse shifted delta}, we have 
        \begin{equation}\label{subset of mixed states: subsequence}
            \tr{\varphi_k}^2 \ \leq \ \dfrac{1+\mu(\theta^{-n_k-1}(\omega))}{2} \leq 1 - \dfrac{1}{2n_0}
        \end{equation}
    and
        \begin{equation}\label{inverse shifted delta:subsequence}
            \sum_{m=1}^d \delta_\omega^{(m)}\left(\varphi_k\right) \ \leq \ 2n_0 \sum_{m=1}^d\ \mbE_{{\phi(\theta^{-n_k-1}(\omega))};\theta^{-n_k-1}(\omega)}\left[\delta^{(m)}_{\omega}\left(\rho^{(n_k)}_{\phi(\theta^{-n_k-1}(\omega));\theta^{-n_k-1}(\omega)}\right)\right]
        \end{equation}
    for all $k\in\mbN$. Note that $(\varphi_k)_{k\in\mbN}$ lies in the compact set
    \begin{equation}\label{subset of mixed}
                \left\{\varrho: \tr{\varrho}^2 \ \leq \ 1 - \dfrac{1}{2n_0}\right\},
    \end{equation}
    so in particular $(\phi_k)_{k\in\mbN}$ has a cluster point $\psi$ satisfying $\tr{\psi}^2 \leq 1 - 1/(2n_0)$, so $\psi$ is a mixed state. By continuity of $\delta^{(m)}$ it follows from \eqref{inverse shifted delta:subsequence} and \eqref{sumbale imlies 0 limit} that
    \begin{equation}
            \sum_{m=1}^d \delta_\omega^{(m)}(\psi) \  = \  0 \ . 
    \end{equation}
\end{proof}

\begin{cor}\label{Weaker version of main theorem}
    For almost every $\omega\in\Omega\setminus X$, there is a mixed state $\psi$ such that for all $a\in\mcA$ there is some $\lambda_a\geq 0$ such that $V_{a;\omega}\psi V^\dagger_{a;\omega} \sim \lambda_a \psi$. 
\end{cor}
\begin{proof}
    The previous proposition gives a mixed state $\psi$ satisfying $\delta_\omega^{(m)}(\psi) = 0$ for all $m\in\{1, \dots, d\}$ for almost every $\omega\in\Omega\setminus X$. Fix such an $\omega\in\Omega\setminus X$ and corresponding $\psi$. Arguing as in \cite[proof of Lemma 3]{maassen2006purification}, $\delta_\omega^{(m)}(\psi) = 0$ for all $m\in\{1, \dots, d\}$ is seen to be equivalent to $\psi$ satisfying, for all $a\in\mcA$, that either $V_{a;\omega}\psi V^\dagger_{a;\omega} = 0$, or else
    \begin{equation}
        \text{
         $\operatorname{tr}\left(\left(V_{a;\omega}\cdot \psi\right)^m\right) \  = \   \operatorname{tr}(\psi^m)$ for all $m\in\{1, \dots, d\}$
        }
    \end{equation}
    In either case, we have that $V_{a; \omega}\psi V^\dagger_{a;\omega} \sim \lambda_a \psi$ for $\lambda_a = \operatorname{tr}\left(V_{a; \omega}\psi V^\dagger_{a;\omega} \right)\geq 0$. 
\end{proof}
\begin{cor}\label{Weaker version of main theorem with dark projections}
    For almost every $\omega\in\Omega\setminus X$, there is a projection $p$ of rank at least 2 for which $p V_{a;\omega}^\dagger V_{a;\omega}p \propto p$ for all $a\in\mcA$. 
\end{cor}
\begin{proof}
    Let $\omega\in\Omega\setminus X$ and $\psi\in\mbS_d$ be as in Corollary \ref{Weaker version of main theorem}. Let $p$ be the range projection of $\psi$. By \cite[Lemma 4]{maassen2006purification}, $p$ satisfies the requirements of the corollary. 
\end{proof}
    \subsubsection{Proof of Proposition \ref{Main theorem, technical part}}
    Corollary \ref{Weaker version of main theorem with dark projections} establishes that $\scrG^{(1)}_\omega$ is nonempty for almost every $\omega\in\Omega\setminus X$. We recall, however, that Proposition \ref{Main theorem, technical part} asserts that $\scrG^{(N)}_\omega$ is nonempty \textit{for all $N\in\mbN$} and almost every $\omega\in\Omega\setminus X$. In this section, we outline how the above proof of Corollary \ref{Weaker version of main theorem} can be modified to prove Proposition \ref{Main theorem, technical part}. 

    The key idea is to ``coarse grain'' in time: we replace the one-step single measurement by the measurement of $N$ outcomes and replace the set of outcomes by finite sequences of outcomes $\mcA^N$.  More formally, for any $N\in\mbN$, let $\mcV^{(N)}$ denote the disordered perfect Kraus measurement  $\left\{V_{\bar{a};\omega}^{(N)}\right\}_{\bar{a}\in\mcA^N}$ on $\mcA^N$. Recall that we write $X_\mcV$ for the purification event corresponding to a measurement ensemble $\mcV$.
    \begin{lemma}\label{Lem:X almost surely equal to XN}
        For all $N\in\mbN$, the almost sure equality $X_\mcV = X_{\mcV^{(N)}}$ holds. 
    \end{lemma}
    \begin{proof}
    In light of the expression of $X_\mcV$ as the intersection 
    \begin{equation}
    X_\mcV
    \  = \  
    \bigcap_{\varrho\in\mbS_d}
    \bigcap_{m\in\mbN}
    \left\{
        \omega
            \,:\,
        \avg_{\mbQ_{\varrho;\omega}}\left[M^{(m,\infty)}_{\varrho;\omega}\right] = 1
    \right\}. 
    \end{equation}
    it suffices to show the $\Pr$-almost sure equality
    \begin{equation}\label{Proof main theorem technical part, Eqn 1}
    \avg_{\mbQ_{\varrho;\omega}}\left[M^{(m,\infty)}_{\varrho;\omega}\right] 
        \  = \ 
    \avg_{\mbQ_{\varrho;\omega; N}}\left[M^{(m,\infty)}_{\varrho;\omega; N}\right]
    \end{equation}
    for all $m\in\mbN$ and $\varrho\in\mbS_d$, 
    where $\mbQ_{\varrho; \omega; N}$ denotes the quantum probability associated to $\mcV^{(N)}_\omega$ and $M_{\varrho; \omega; N}^{(m, \infty)}$ denotes the quantity \eqref{Submartingale limit function definition} corresponding to $\mcV^{(N)}_\omega$. We also write $M_{\varrho; \omega; N}^{(m, n)}$ denote the quantity \eqref{Submartingale definition} for $\mcV^{(N)}_\omega$, so that $M_{\varrho; \omega; N}^{(m, \infty)} = \lim_{n\to\infty} M_{\varrho; \omega; N}^{(m, n)}$ holds with $\mbQ_{\varrho; \omega}$-probability 1. 
    
    To prove \eqref{Proof main theorem technical part, Eqn 1}, note that $M_{\varrho; \omega; N}^{(m, n)}(\bar{a}) = M_{\varrho; \omega}^{(m, nN)}(\bar{a})$ for any $\bar{a}\in\mcA^{nN}$ and $n\in\mbN$. Thus, the dominated convergence theorem and the fact that $\left(M_{\varrho; \omega}^{(m, n)}\right)_{n\in\mbN}$ is a submartingale imply that
    \begin{align}
     \avg_{\mbQ_{\varrho;\omega}}\left[M^{(m,\infty)}_{\varrho;\omega}\right] 
     &\  = \  
     \lim_{n\to\infty}
    \avg_{\mbQ_{\varrho;\omega}}\left[M^{(m,n)}_{\varrho;\omega}\right] 
    \\
    &\  = \  
     \lim_{n\to\infty}
    \sum_{\bar{a}\in\mcA^n}
    \mbQ_{\varrho; \omega}^{(n)}(\bar{a})\operatorname{tr}\left(\rho^{(n)}_{\varrho; \omega}(\bar{a})^m\right)
    \\
    &\  = \  
     \lim_{n\to\infty}
    \sum_{\bar{a}\in\mcA^{nN}}
    \mbQ_{\varrho; \omega}^{(nN)}(\bar{a})\operatorname{tr}\left(\rho^{(nN)}_{\varrho; \omega}(\bar{a})^m\right)
    \\
    &\  = \  
     \lim_{n\to\infty}
    \avg_{\mbQ_{\varrho;\omega; N}}\left[M^{(m,n)}_{\varrho;\omega; N}\right] 
    \\
    &\  = \  
    \avg_{\mbQ_{\varrho;\omega; N}}\left[M^{(m,\infty)}_{\varrho;\omega; N}\right], 
    \end{align}
    as desired. Thus, $X_\mcV = X_{\mcV^{(N)}}$, as desired. 
    \end{proof}
    In particular, the above lemma says that a property $\mathfrak{P}$ holds for almost every $\omega\in X_\mcV$ if and only if $\mathfrak{P}$ holds for almost every $\omega\in X_{\mcV^{(N)}}$.
    \begin{proof}[Proof of Proposition \ref{Main theorem, technical part}]
    Observe that in sections \ref{Subsec:Submartingale incremenets} and \ref{Subsec:Measurability considerations}, we only used that $\mcV$ was a disordered perfect Kraus measurement ensemble on a finite outcome space, that $\theta$ was measure-preserving and invertible, and that $\Pr(X_\mcV)\in\{0, 1\}$. Lemma \ref{Lem:X almost surely equal to XN} and Theorem \ref{0-1 law for purification of disordered quantum trajectories} imply that $\Pr(X_{\mcV^{(N)}})\in\{0, 1\}$ for any $N\in\mbN$. Thus, by replacing $\mcA$ with $\mcA^N$, $\mcV$ with $\mcV^{(N)}$, and $\theta$ with $\theta^{N}$ in Corollary \ref{Weaker version of main theorem with dark projections} we find that for almost every $\omega\in \Omega\setminus X_{\mcV^{(N)}}$, there is a projection $p$ of rank at least 2 for which 
    \begin{equation}\label{Proof of Proposition main theorem technical part, Eqn 1}
       p V_{\bar{a};\omega}^{(N)\dagger} V_{\bar{a};\omega}^{(N)}p \ \propto \ p \quad \text{for all $\bar{a}\in\mcA^N$ .}
    \end{equation}
    %
    %
    By Definition \ref{Dark subspaces: n-dark projections of rank r definition} and one more application of Lemma \ref{Lem:X almost surely equal to XN}, we see that $p\in\scrG^{(N)}_\omega$ for almost every $\omega\in\Omega\setminus X_\mcV$, which is precisely what Proposition \ref{Main theorem, technical part} asserts. 
    \end{proof}
\subsection{Proof of Theorem \ref{Main theorem'}}
We are now ready to prove Theorem \ref{Main theorem'}, restated here for the reader's convenience:
\thirdresult*
\begin{proof}[Proof of Theorem \ref{Main theorem'}]
    The measurability of $\omega \mapsto \scrD_\omega$ was established in Corollary \ref{Dark subspaces: Gray and dark subspaces are random sets}. 
    
    For the other part, note that, by Proposition \ref{Main theorem, technical part}, the set $\scrG^{(N)}_\omega$ is nonempty for all $N\in\mbN$ and almost every $\omega\in\Omega\setminus X_\mcV$. So, by Corollary \ref{Dark subspaces: Sufficient condition for dark subspaces nonempty}, $\scrD_\omega\neq\emptyset$ a.s. on $\Omega\setminus X_{\mcV}$. Thus, the almost sure inclusion $X_\mcV \supset \{\scrD = \emptyset\}$ holds. On the other hand, for almost every $\omega\in\Omega$ such that $\scrD_\omega\neq\emptyset$, there is a projection $p$ of rank $r\geq 2$ such that for all $n\in\mbN$ and $\bar{a}\in\mcA^n$, $\lambda_{\bar{a}}^{(n)} = \mbQ_{\varrho; \omega}(\bar{a})$ satisfies $p V_{\bar{a}; \omega}^{(n)\dagger} V_{\bar{a}; \omega}^{(n)} p = \lambda_{\bar{a}}^{(n)} p$. So, if we let $\varrho = p/r\in\mbS_d$, we find that, $\mbQ_{\varrho; \omega}$-almost surely,
    \begin{align}
        \operatorname{tr}
        \left(
            \rho^{(n)}_{\varrho; \omega}(\bar{a})^2
        \right)
        &\  = \  
        \operatorname{tr}
        \left(
            \cfrac{V_{\bar{a}; \omega}^{(n)}\varrho 
            V_{\bar{a}; \omega}^{(n)\dagger}V_{\bar{a}; \omega}^{(n)}\varrho 
            V_{\bar{a}; \omega}^{(n)\dagger}}{\left[\tr{V_{\bar{a}; \omega}^{(n)}\varrho 
            V_{\bar{a}; \omega}^{(n)\dagger}}\right]^2}
        \right)
        \\
        &\  = \  
        \operatorname{tr}
        \left(
            \cfrac{V_{\bar{a}; \omega}^{(n)}p 
            V_{\bar{a}; \omega}^{(n)\dagger}V_{\bar{a}; \omega}^{(n)}p 
            V_{\bar{a}; \omega}^{(n)\dagger}}
            {\left[\tr{
             p V_{\bar{a}; \omega}^{(n)\dagger}V_{\bar{a}; \omega}^{(n)}p 
             }\right]^2}
        \right)
        \\
        &\  = \  
        \operatorname{tr}
        \left(
            \cfrac
            {
            \left(
                p V_{\bar{a}; \omega}^{(n)\dagger}V_{\bar{a}; \omega}^{(n)}p 
            \right)^2
            }
            {
             \left[\tr{
             p V_{\bar{a}; \omega}^{(n)\dagger}V_{\bar{a}; \omega}^{(n)}p 
             }\right]^2
            }
        \right)
        \\
        &\  = \  
        1/r \ .
    \end{align}
    In particular, $\lim_{n\to\infty} \operatorname{tr}
        \left(
            \rho^{(n)}_{\varrho; \omega}(\bar{a})^2
        \right) = 1/r < 1$ for quantum almost every $\bar{a}\in\mcA^\mbN$, implying that $\omega\in\Omega\setminus X_\mcV$ $\Pr$-almost surely. Thus, the almost sure inclusion $X_\mcV\subset \{\scrD = \emptyset\}$ holds. 

    Lastly, because Theorem \ref{0-1 law for purification of disordered quantum trajectories} implies that $\Pr(X_\mcV)\in\{0, 1\}$, the almost sure equality $X_\mcV = \{\scrD = \emptyset\}$ gives that $\mcV$ asymptotically purifies if and only if $\Pr(\scrD = \emptyset) > 0$. 
\end{proof}
%
%


%
%
\section*{Acknowledgements}
 This material is based upon work supported by the National Science Foundation under Grant No. 2153946.
\appendix 
\section{Appendix}\label{App:Measurability considerations}
   \subsection{Measurability}
\begin{lemma}\label{Gamma is measurable}
    For each $n\in \mbN$, the $\mbM_d$-valued map on $\mbS_d\times\mcA^n \times \Omega$ defined in \eqref{Measurability of Gamma map}, i.e.,
   \begin{equation}  \Gamma(\varrho, \bar{a},\omega) \ := \ \rho^{(n)}_{\varrho;\omega}(\bar{a}) , \end{equation}
    is $(\mcB(\mbS_d)\otimes\Sigma_n\otimes \mcF)$-measurable for each $n\in\mbN$. 
\end{lemma}
\begin{proof}
This follows directly due to the fact that $\Gamma$ is a Carathéodory function in the sense that for each $(\bar{a},\omega)$, $\varrho\mapsto \Gamma(\varrho, \bar{a},\Omega)$ is continuous in $\varrho$ and for each $\varrho$ the map $(\bar{a},\omega)\mapsto \Gamma(\varrho, \bar{a},\Omega)$ is $\Sigma\times\mcF$-measurable. The joint measurability of $\Gamma$ follows from the fact that Carathéodory functions are jointly measurable \cite[Lemma 4.51]{guide2006infinite}. 
\end{proof}

Recall that, by Tychonoff's theorem, $\mcA^\mbN$ is a compact topological space when given the product topology coming from the discrete topology on $\mcA$. Furthermore $\mcA^\mbN$ is a  metrizable  space. Explicitly, let $d(\bar{a}, \bar{b})$ by given by 
\begin{equation}
    \ell(\bar{a}, \bar{b})
            =
        \sum_{n\in\mbN}
        2^{-n}
        \cdot 
        \delta_{\pi_n(\bar{a})\neq \pi_n(\bar{b})} \ .
\end{equation}
Clearly, $\ell(\bar{a}, \bar{b}) = 0$ if and only if $\bar{a} = \bar{b}$. Moreover, because $\delta_{\pi_n(\bar{a})\neq \pi_n(\bar{b})}$ satisfies the triangle inequality,  $\ell$ is a metric on $\mcA^\mbN$. It is a straightforward exercise to check that the metric topology generate by $\ell$ is product topology. Furthermore, the topology is separable, since
\begin{equation}
    \bigcup_{n\in\mbN}\bigcup_{a_1, \dots, a_n\in\mcA}\{\pi_n^{-1}(a_1, \dots, a_n)\} \ .
\end{equation}
is a base.

If we let $\mcP(\mcA^\mbN)$ denote the set of probability measures on $\mcA^\mbN$ and equip $\mcP(\mcA^\mbN)$ with the Prokhorov metric
\begin{equation}
     \pi(\mu, \nu)
    =
    \inf\Big\{
        \varepsilon>0
            \,\,:\,\,
        \text{$\mu(A) \leq \nu(A^\varepsilon) + \varepsilon$
        and $\nu(A) \leq \mu(A^\varepsilon) + \varepsilon$ for all $A\in\mcB(\mcA^\mbN)$}
                \Big\}\ ,
\end{equation}
where, for $A\in\mcB(\mcA^\mbN)$, $A^\varepsilon$ denotes the set 
\begin{equation}
    A^\varepsilon = \left \{ \bar{b} \ : \ \inf_{\bar{a}\in A}\ell(\bar{b},\bar{a}) < \epsilon \right \} \ ,
\end{equation}
then the topology on $\mcP(\mcA^\mbN)$ generated by $\pi$ is the weak$^*$ topology on $\mcP(\mcA^\mbN)$ inherited by viewing $\mcP(\mcA^\mbN)$ as the Banach space dual of the set $\mcC(\mcA^\mbN)$ of continuous functions on $\mcA^\mbN$; see \cite[Apendix III, Theorem 5]{billingsley2013convergence}. 
\begin{lemma}\label{The quantum probability is jointly measurable}
    Let $\{V_a\}_{a\in\mcA}$ be a disordered perfect Kraus measurement on $\mcA$. The map
     \begin{align}
     \begin{split}
         \mbS_d\times\Omega&\to\mcP(\mcA^\mbN)\\
         (\varrho, \omega)&\mapsto \mbQ_{\varrho; \omega}
         \end{split}
    \end{align}
     is $(\borel{\mbS_d}\otimes\mcF)$-measurable when $\mcP(\mcA^\mbN)$ is given the Borel $\sigma$-algebra induced by the Prokhorov metric.
\end{lemma}
\begin{proof}
We have already seen that the Borel $\sigma$-algebra induced by the Prokhorov metric is the same as that generated by the weak$^*$ topology, so it suffices to show that, for any continuous real-valued function $f$, on $\mcA^\mbN$ ($f\in\scrC(\mcA^\mbN)$), the map 
\begin{equation}
    (\varrho, \omega)\ \mapsto \
    \mbE_{\mbQ_{\varrho; \omega}}(f)
\end{equation}
is measurable. 

Because $\mcA^\mbN$ is a compact space, all $f\in\scrC\left(\mcA^\mbN\right)$ are bounded. We claim that any $f\in \scrC\left(\mcA^\mbN\right)$ is a uniform limit of functions of the form $f_n\circ \pi_n$ with $f_n\in \scrC\left (\mcA^n \right )$.  This could be obtained from the Stone-Weierstrass theorem, but can also be seen directly by defining $f_n = f \circ \pi_n^{(a)}$ where $\pi_n^{(a)}(a_1, \dots, a_n, a_{n+1})=(a_1, \dots, a_n, a, \dots)$, with $a$ an arbitrary fixed element of $\mcA$. The dominated convergence theorem then implies that
\begin{equation}
    \mbE_{\mbQ_{\varrho; \omega}}
    \left(
    f
    \right)
    \ = \
    \lim_n \mbE_{\mbQ_{\varrho; \omega}}
    \left(
    f_n
    \right) \ ,
\end{equation}
hence it suffices to consider $f$ of the form $g\circ \pi_n$, i.e. $f$ that depends on only finitely many variables. For such $f$, t 
\begin{align}
    \mbE_{\mbQ_{\varrho; \omega}}(f)
    \ &= \
    \mbE_{\mbQ^{(n)}_{\varrho; \omega}}(g)\\
    \ &= \ 
    \sum_{\bar{a}\in\mcA^n}
    \operatorname{tr}\left(V_{\bar{a}; \omega}^{(n)}\varrho V_{\bar{a}; \omega}^{(n)\dagger}\right)g(\bar{a}) \ ,
\end{align}
which is clearly measurable in $(\varrho, \omega)$.
\end{proof}

The following lemma shows that a re-indexing of a sequence of measurable functions by a $\mbN$-valued (or  $\mbN\cup\{\infty\}$-valued) measurable map is also measurable. The proof is similar the proof that $X_\tau$ is measurable where $(X_t)_{t\in T}$ is a stochastic process and $\tau$ is a $T$-valued stopping time. For a sequence of measurable maps $(X_n)_{n\in\mbN}$ and a $\mbN\cup\{\infty\}$-value measurable function, $\tau$, we shall call $X_\tau$ the stopped random variable or the re-indexing of $(X_n)_{n\in\mbN}$ with respect to $\tau$.

\begin{lemma}\label{stopped random elememt} 
Let $(X_n)_{n\in\mbN}$ be a sequence of measurable maps defined on a measurable space $(X,\mcX)$ taking values on a measurable space $(U,\mcY)$. If $\tau:X \to \mbN\cup\{\infty\}$ be an $\mcX$-measurable random map (so that $\{\tau=n\}$ and $\{\tau=\infty\}$ are elements in $\mcX$), then the map $X_\tau$ defined by  
\begin{equation}
    X_\tau(x) = \begin{cases}
        X_{\tau(x)}(x) & \text{ if } \tau(x) < \infty\\
        y_0 & \text{ if } \tau(x) = \infty
    \end{cases}
\end{equation}
for some fixed $y_0\in Y$ is also a measurable map. 
\end{lemma}
\begin{proof}
    Let $B\in\mcY$ then we have that 
    \begin{equation}
        \{x : X_\tau \in B\} = \{x: X_\tau(x) \in B \text{ and } \tau(x)<\infty\} \sqcup \{x: X_\tau(x) \in B \text{ and } \tau(x) =\infty\} \ .
    \end{equation}
First note that $\{x : X_\tau(x) \in B \text{ and } \tau(x) =\infty\} $ is either $\{x :\tau(x) = \infty\}$ (if $y_0\in B$) or $\emptyset$ (if $y_0\not \in B$); in either case it is measurable. On the other hand we have  
\begin{align}
    \{x: X_\tau(x) \in B \text{ and } \tau(x)<\infty\} &= \bigcup_{n\in\mbN}\{x: X_\tau(x) \in B \text{ and } \tau(x)=n\}\\
    &= \bigcup_{n\in\mbN}\left(\{x: X_n(x) \in B \} \cap\{x: \tau(x)=n\}\right) \ .
\end{align}
Therefore we have that $X_\tau$ is indeed $(\mcY,\mcX)$-measurable. 
\end{proof}
\subsection{Equicontinuity}\label{App:Equicontinuity}
In this section, we prove Proposition \ref{Equicontinuity}: 
\Equicontinuity*
\noindent Recall that the maps $\Phi^{(n)}_\omega:\mbS_d\to [0,1]$, as defined in \eqref{Uppercase phi},  are given by:
$$    \Phi_\omega^{(n)}(\varrho )\ = \ \mbE_{\varrho;\omega}\left[M_{\varrho;\omega}^{(2, n)}(\bar{a})\right] \ . $$ 
\begin{proof}[Proof of Proposition \ref{Equicontinuity}]
It is enough to prove that each $\Phi_\omega^{(n)}$ is $K$-Lipschitz, where $K\geq0$ is uniform in $n$. To simplify notation, we write $F_n$ for $\Phi_\omega^{(n)}$ and, for any $\bar{a}\in\mcA^n$, write $F_{\bar{a}}$ to denote the function
\begin{equation}
    \varrho\ \mapsto \ \tr{V_{\bar{a}; \omega}^{(n)} \varrho V_{\bar{a}; \omega}^{(n)\dagger}}\tr{\left(V_{\bar{a}; \omega}^{(n)} \cdot \varrho\right)^2} \ ,
\end{equation}
so $F_n = \sum_{\bar{a}\in\mcA^n}F_{\bar{a}}$. Note that $F_{\bar{a}}\geq 0$. Letting
\begin{equation}
    g_{\bar{a}}(\varrho)
        \ := \
    \tr{\left(V_{\bar{a}; \omega}^{(n)}\varrho V_{\bar{a}; \omega}^{(n)\dagger}\right)^2}
        \quad\text{and}\quad 
    h_{\bar{a}}(\varrho)
       \ := \
    \tr{V_{\bar{a}; \omega}^{(n)}\varrho V_{\bar{a}; \omega}^{(n)\dagger}} \ .
\end{equation}
it is clear that $g_{\bar{a}}, h_{\bar{a}}\geq 0$ and
\begin{equation}
    F_{\bar{a}}
   \ = \
    \begin{cases}
        \cfrac{g_{\bar{a}} 
        }{
        h_{\bar{a}}
        } &\text{whenever }h_{\bar{a}}> 0\\
        0 &\text{else} \ .
    \end{cases}
\end{equation}
Since
\begin{align}
    \cfrac{g_{\bar{a}}(\varrho) 
        }{
        h_{\bar{a}}(\varrho)
        }
        \ &= \
        \tr{\left(\cfrac{V_{\bar{a}; \omega}^{(n)}\varrho V_{\bar{a}; \omega}^{(n)\dagger}}{\tr{V_{\bar{a}; \omega}^{(n)}\varrho V_{\bar{a}; \omega}^{(n)\dagger}}}\right)V_{\bar{a}; \omega}^{(n)}\varrho V_{\bar{a}; \omega}^{(n)\dagger}}\\
        &= \
        \tr{\varrho^{1/2}V_{\bar{a}; \omega}^{(n)\dagger}\left(\cfrac{V_{\bar{a}; \omega}^{(n)}\varrho V_{\bar{a}; \omega}^{(n)\dagger}}{\tr{V_{\bar{a}; \omega}^{(n)}\varrho V_{\bar{a}; \omega}^{(n)\dagger}}}\right)V_{\bar{a}; \omega}^{(n)}\varrho^{1/2}}
        \ \leq  \
        h_{\bar{a}}(\varrho) \ ,\label{Equicontinuity, Eqn 1 (Bound on F bar a)}
\end{align}
whenever $h_{\bar{a}}(\varrho) > 0$.  Here, we use that if $X\geq 0$ is nonzero, then
$ \frac{X}{\tr{X}}\leq \frac{X}{\|X\|}\leq I $. It follows that the functions $F_{\bar{a}}$ are continuous, and thus $F_n$ is continuous also.\footnote{Even more,
\[
    F_n(\varrho)
    \ = \
    \sum_{\bar{a}\in\mcA^n}
    F_{\bar{a}}(\varrho)
    \ \leq \ 
    \sum_{\bar{a}\in\mcA^n}
    \tr{V_{\bar{a}; \omega}^{(n)}\varrho V_{\bar{a}; \omega}^{(n)\dagger}}
   \ = \ 
    1 \ ,
\]
so $F_{n}$ is bounded uniformly in $n$.}

Now fix $\varrho, \varrho'\in\mbS_d$ and consider the function 
\begin{align}\label{Equicontinuity, Eqn 2 (Line between rho and rho')}
    \begin{split}
    \varrho_{(\cdot)}:[0, 1] &\to \mbS_d\\
    \varrho_t &:= t\varrho + (1 - t)\varrho' \ .
    \end{split}
\end{align}
We begin by proving the map 
\begin{equation}
    t\mapsto F_n(\varrho_t)
\end{equation}
is differentiable on $(0, 1)$. Because $F_n = \sum_{\bar{a}\in\mcA^n}F_{\bar{a}}$, it suffices to prove that the map
\begin{align}
    \begin{split}
    G_{\bar{a}}: [0, 1] &\to \mbR\\
    t& \ \mapsto \  F_{\bar{a}}(\varrho_t) \ = \ \frac{g_{\bar{a}}(\varrho_t)}{h_{\bar{a}}(\varrho_t)} 
    \end{split}
\end{align}
is differentiable on $(0, 1)$ for each $\bar{a}\in\mcA^n$. Fix $\bar{a}\in\mcA^n$. Since $h_{\bar{a}}(\varrho_t)$ and $g_{\bar{a}}(\varrho_t)$ are both clearly differentiable, $G_{\bar{a}}$ is seen to be differentiable whenever $h_{\bar{a}}(\varrho_t)\neq 0$. On the other hand, if $h_{\bar{a}}(\varrho_t)=0$ then $F_{\bar{a}}(\varrho) = F_{\bar{a}}(\varrho') = 0$, and  $G_{\bar{a}}\equiv 0$, so differentiability is clear. Thus we see that $G_{\bar{a}}$ is differentiable for $t\in (0,1)$ and furthermore $h_{\bar{a}}(\varrho_t)$ is either identically zero for all $t\in (0,1)$ or non-zero for all $t\in (0,1)$.

Now, for any $t\in (0,1)$, we have 
\begin{equation}\label{Equicontinuity, Eqn 3 (Derivative of G_n(t))}
    F_n'(\varrho_t)
    \ = \ 
    \sum_{\bar{a}\in C}G'_{\bar{a}}(t)
\end{equation}
where 
\begin{equation} C \ = \ \left\{
    \bar{a}\,\,:\,\,
    h_{\bar{a}}(\varrho_t) > 0 \text{ for }t\in (0,1)
\right\} \ .
\end{equation}
For $\bar{a}\in C$, we compute 
\begin{align}
    \left|
        G'_{\bar{a}}(t)
    \right|
   \ &\leq  \
     \left|
        \frac{g'_{\bar{a}}(\varrho_t)}{h_{\bar{a}}(\varrho_t)}
     \right|
     +
    \left|
        \frac{ g_{\bar{a}}(\varrho_t) h'_{\bar{a}}(\varrho_t)}{h_{\bar{a}}(\varrho_t)^2}
    \right|\\
    &= \
     \left|
        \frac{
        2\tr{
            V_{\bar{a};\omega}^{(n)}
            \varrho_t
            V_{\bar{a};\omega}^{(n)\dagger}
            V_{\bar{a};\omega}^{(n)}
            (\varrho - \varrho')
            V_{\bar{a};\omega}^{(n)\dagger}
            }
        }{h_{\bar{a}}(\varrho_t)}
     \right|
     +
    \left|
    \frac{
    g_{\bar{a}}(\varrho_t)
    }
    {
    h_{\bar{a}}(\varrho_t)^2
    }    \tr{V_{\bar{a};\omega}^{(n)}
            (\varrho - \varrho')
            V_{\bar{a};\omega}^{(n)\dagger}}
    \right|\\
    &\leq 
        2\tr{
            \left(\cfrac{V_{\bar{a};\omega}^{(n)}
            \varrho_t
            V_{\bar{a};\omega}^{(n)\dagger}}{\tr{V_{\bar{a};\omega}^{(n)}
            \varrho_t
            V_{\bar{a};\omega}^{(n)\dagger}}}\right)
            V_{\bar{a};\omega}^{(n)}
            |\varrho - \varrho'|
            V_{\bar{a};\omega}^{(n)\dagger}
            }\\
    &\hspace{15mm}
     +
     \tr{\left(
     \cfrac{
     V_{\bar{a};\omega}^{(n)}
            \varrho_t
            V_{\bar{a};\omega}^{(n)\dagger}
     }
     {
     \tr{V_{\bar{a};\omega}^{(n)}
            \varrho_t
            V_{\bar{a};\omega}^{(n)\dagger}}
     }
     \right)^2
     }
     \tr{V_{\bar{a};\omega}^{(n)}
            |\varrho - \varrho'|
            V_{\bar{a};\omega}^{(n)\dagger}}\\
    &\leq \
    (2 + d)
    \tr{V_{\bar{a};\omega}^{(n)}
            |\varrho - \varrho'|
            V_{\bar{a};\omega}^{(n)\dagger}} \ .
\end{align}
Thus, 
\begin{align}
    \abs{F_n'(t)} 
     \ &\leq \ 
    (2 + d)\sum_{\bar{a}\in\mcA^n}
    \tr{V_{\bar{a};\omega}^{(n)}\abs{\varrho - \varrho'}V_{\bar{a};\omega}^{(n)\dagger}}\notag\\
    \ &= \
     (2 + d)\operatorname{tr}\abs{\varrho - \varrho'}\label{Equicontinuity, Eqn 6 (Bound on derivative of G)}
\end{align}
for any $t\in (0, 1)$. By the mean value theorem, we conclude that 
\begin{equation}
     \abs{F_n(\varrho) - F_n(\varrho')}
    \ \leq \
    (2 + d)\operatorname{tr}\abs{\varrho - \varrho'} \ .
\end{equation}

Because this bound is independent of $\varrho, \varrho'$, we see that $F_n$ is $(2 + d)$-Lipschitz continuous uniformly in $n$, which implies the desired uniform equicontinuity. 
\end{proof}

 
%
%

\printbibliography

\end{document}